\def\gsim{\mathrel{\scriptstyle{\buildrel > \over \sim}}}
\def\lsim{\mathrel{\scriptstyle{\buildrel < \over \sim}}}
\magnification 1200
\baselineskip=17pt

\centerline{\bf THEORY OF DECOUPLING IN THE MIXED PHASE OF}
\bigskip
\centerline{\bf EXTREMELY TYPE-II LAYERED SUPERCONDUCTORS}
\vskip 50pt
\centerline{J. P. Rodriguez}
\medskip
\centerline{\it Dept. of Physics and Astronomy,
California State University, Los Angeles, CA 90032.}
\vskip 30pt
\centerline  {\bf  Abstract}
\vskip 8pt\noindent
The mixed phase of extremely type-II layered superconductors
in perpendicular magnetic field 
is studied theoretically via the layered $XY$ model with
uniform frustration.  A partial duality analysis is carried
out in the weak-coupling limit.  It consistently accounts
for both intra-layer (pancake) and inter-layer
(Josephson) vortex excitations.  The main conclusion reached
is that dislocations  of the two-dimensional (2D)
vortex lattices within layers 
drive a unique second-order melting transition
at high perpendicular fields
between a low-temperature superconducting phase that displays
a Josephson effect and a high-temperature ``normal'' phase
that displays no Josephson effect.  
The former state is best described by weakly
coupled 2D vortex lattices, while the latter state is
best characterized by a decoupled vortex liquid.
It is further argued on the basis of the duality
analysis that the second-order melting transition
converts itself into a first-order one as the
perpendicular field is lowered and approaches
the dimensional cross-over scale.  The resulting
critical endpoint potentially accounts for the
same phenomenon  that is observed in the mixed phase
of clean high-temperature superconductors.

\bigskip
\noindent
PACS Indices:   74.60.-w, 74.25.Dw, 74.25.Ha, 74.60.Ge

\vfill\eject
\centerline{\bf I. Introduction}
\bigskip

The nature of the thermodynamic phase diagram in clean high-temperature
superconductors in external magnetic field has been elucidated only
recently.$^1$  Experimental evidence for 
a first-order melting transition of the
Abrikosov vortex lattice has  been obtained from the      
measurement of a jump in the magnetization and from
the measurement of a   latent heat.$^{2,3}$  
These two quantities are found  to 
satisfy the Clausius-Clapeyron relation, which indicates
that the first-order transition is indeed thermodynamic.
Although the existence of such a melting transition was expected
on theoretical grounds due to the extreme type-II and anisotropic
character of high-temperature superconductors,$^4$ the
multi-critical phenomenon
 exhibited by the associated phase diagram for the mixed phase
was  not.  In particular, the first-order melting line begins at
(or near) the zero-field transition, but it ends strangely
in the middle of the field versus temperature 
($T$ - $H_{\perp}$) plane.  The critical end point, furthermore,
coincides (or is near) a multi-critical point, from which a
nearly field-independent
``vertical'' depinning line of the
 vortex lattice begins at higher fields.$^{5,6}$
A  weakly temperature-dependent 
``horizontal'' second-peak line also continues on from the multi-critical
point to lower temperatures.  Bulk pinning becomes appreciably stronger
at fields above this line.$^7$

The first-order    melting observed in clean high-temperature
superconductors is commonly interpreted theoretically in terms
of elastic medium descriptions of the vortex lattice.$^4$
The melting point  in such theories is usually determined
by the Lindemann criterion, however, which cannot predict
the nature of the transition.  In other words, the  theory
is compatible with either   a first-order,
a second-order, or even a cross-over transition.
Elastic medium theory has also been used to describe the mixed
phase of layered superconductors in particular.$^8$  A 
first-order decoupling transition is predicted
to occur  in the extreme type-II
limit, beyond which absolutely no
Josephson coupling remains in between the layers.$^9$
A subsequent high-temperature
analysis of the anisotropic $XY$ model with uniform frustration
has demonstrated, however, that a small degree of local Josephson
coupling must survive the decoupling transition.$^{10,11}$
Last, it is important to remark that  most elastic medium calculations
to date have been performed in the gaussian approximation.$^4$
They therefore  neglect important
topological excitations of the vortex lattice,
such as dislocations, that can drive the melting
transition.$^{12,13}$

Other more numerical approaches to the theoretical description of
the mixed phase in clean high-temperature superconductors
have succeeded in obtaining a clear first-order melting
transition.  These include Monte-Carlo simulations of both the
frustrated $XY$ model$^{14,15}$ and of
Ginzburg-Landau theory.$^{16}$  The former
simulations  are plagued, however, by
equilibration problems at low temperatures
 due to the artificial pinning 
caused     by the underlying $XY$  model lattice.$^{11}$
The latter simulations of Ginzburg-Landau theory,
on the other hand, are 
performed in the lowest-Landau-level approximation.
This approximation  is in principle valid only in the
vicinity of the mean-field $H_{c2}$ transition,
which remains far from the vortex-lattice melting
line observed experimentally in high-temperature
superconductors because of anisotropy.$^{2,3}$

In this paper, we shall develop a theory for decoupling
in the mixed phase of extreme type-II layered superconductors
that is based on an analysis of the corresponding layered $XY$
model with uniform frustration.  The analysis is effected
through a {\it partial} duality transformation
that leads to a useful neutral
Coulomb-gas (CG) description for the Josephson coupling.$^{17-20}$
Unlike some of  the Coulomb-gas analyzes of
the layered  $XY$ model that have been performed 
in the past,$^{21-24}$
both gaussian and topological excitations of the vortex lattice
within layers 
are {\it consistently} accounted for  
in this approach.
This consistency is responsible for the primary result of the paper,
which is that there exists a unique second-order transition in the
weak-coupling limit that separates a coupled superconducting phase 
at low temperature
from a decoupled ``normal'' state at high temperature.
In  the context of the mixed phase,
the weak-coupling limit is reached 
at field components perpendicular to the layers that are
large compared to the dimensional cross-over scale
$B_{\perp}^* = \Phi_0/\Lambda_0^2$ (see Fig. 1).  
Here, $\Lambda_0$ denotes the Josephson penetration length.
Note that the criterion used throughout to 
discriminate between a coupled and a
decoupled state is the presence or absence of a macroscopic Josephson
effect, respectively.  
The above result implies that
neither the Friedel scenario,$^{4,25}$
 which is a state composed of 
decoupled superconducting layers, nor a ``line-liquid'' state,$^{14,24,26}$
which we define here as a set of
 normal layers in external magnetic field
 that exhibit a Josephson effect,
are likely to be thermodynamic states in the mixed phase of
extreme type-II layered
superconductors when disorder is absent.  The prediction of a
unique phase transition in the uniformly frustrated $XY$ model
with anisotropy  is notably consistent with recent extensive
Monte Carlo simulations.$^{15}$
It was also  anticipated
by various authors in studies of the layered $XY$ model
without perpendicular frustration.$^{21, 22, 27, 28}$ 

The duality analysis mentioned above is also employed in the paper
to map out the phase diagram of the layered $XY$ model with uniform
frustration.  
A finite number of layers is assumed.
The weak-coupling limit at  high perpendicular
fields is found to contain a low-temperature
phase  made up of independent two-dimensional (2D) vortex lattices that
sustain a Josephson effect.
This phase is separated from a decoupled liquid of planar vortices at high
temperatures by a second-order melting transition akin to that shown by
an isolated 2D vortex lattice.$^{29-33}$  
The existence of such a second-order melting transition is
demonstrated by making note of a formal equivalence 
between the Coulomb gas
descriptions of the 
layered $XY$ model with and  without frustration.$^{20,34}$ 
Also, the former smectic
2D vortex-lattice phase is likely to be a
type of ``supersolid'' matter (see sect. IV.C and ref.  13).  The above
weak-coupling analysis is then shown to break down at sufficiently
low perpendicular fields 
of order the dimensional cross-over scale $B_{\perp }^*$.
It is argued that this breakdown signals a cross-over transition to
a flux-line lattice regime at low fields and temperatures, while that
it signals a first-order decoupling transition to a potentially
defective flux-line lattice state at low fields but high temperatures.
It is further argued on this basis that the
first-order decoupling transition terminates at a critical endpoint
at a temperature of order the
2D vortex-lattice melting temperature and at a perpendicular field
of order many times the dimensional cross-over scale
 $B_{\perp }^*$.  The resulting phase diagram
(Fig. 1) is strikingly
similar to those reported for the mixed phase of clean high-temperature
superconductors.$^{2,3,5-7}$

The paper is organized in the following way.  Section II contains
the weak-coupling duality analysis of the uniformly frustrated
$XY$ model.   The principal input is the nature of the phase
coherence in isolated layers, which is assumed to be either
quasi long range [Eq. (17)] or short range [Eq. (18)].
The output of the weak-coupling  analysis is that 
the former implies the presence
of a Josephson effect, whereas the latter implies its absence.
Equations (43) and (55) for the local Josephson coupling in the respective
coupled and decoupled phases summarize these results. 
This theory is then  used    to determine
the phase diagram of 
extremely type-II layered superconductors in
perpendicular magnetic field in section III.
The principal new result is the prediction of a coupled 2D vortex lattice
phase at low temperatures and high fields that notably displays
a Josephson plasma resonance with exponential temperature dependence
[see Fig. 1 and Eq. (66)].  
A comparison of these results with previous experimental and theoretical
work is made in section IV,   while  some conclusions are reached in
section V.  Finally, technical issues 
that concern the long-range
nature of phase correlations in the 2D $XY$ model and 
that concern
a fermion analogy employed to describe the Josephson coupling between
layers are treated in Appendix A and B, respectively.

\bigskip
\bigskip
\centerline{\bf II. Theory}
\bigskip
The thermodynamics in the interior of the mixed phase of 
extremely type-II superconductors ($\lambda_{\rm L}\rightarrow\infty$)
can be described at least qualitatively  by the
uniformly frustrated $XY$ model over the cubic lattice.$^{14,15,24}$  
In the case where the superconductor is
composed of $N$  Josephson coupled layers, 
the corresponding kinetic energy for the superfluid flow reads
$$\eqalignno{ E_{XY}^{(3)} = - J_{\parallel}&
\sum_{l=1}^{N}\sum_{\vec r}\sum_{\mu=x,y}
{\rm cos}[\Delta_{\mu}\phi(\vec r,l)-A_{\mu}(\vec r,l)]\cr
& - J_{\perp}\sum_{l=1}^{N-1}\sum_{\vec r}
{\rm cos}[\phi(\vec r, l+1)-\phi(\vec r, l)-A_z(\vec r,l)].
& (1) \cr}$$
Above, $\phi(\vec r, l)$ is the superconducting phase at a point
$\vec r$ in layer $l$, while 
$A_{\mu} = (0, b_{\perp} x, -b_{\parallel}x)$
is the   vector potential.  Also, $\Delta_{\mu}$ denotes the
nearest-neighbor difference operator along the $\hat\mu$ direction 
of  the square lattice.
The magnetic induction
parallel and perpendicular to the layers
is related to the frustration $\vec b$ through the respective
identities $B_{\parallel} = (\Phi_0/2\pi d) b_{\parallel}$
and $B_{\perp} = (\Phi_0/2\pi a) b_{\perp}$.
Here the length 
scales   $d$ and $a$ denote respectively the separation between adjacent 
layers and the lattice constant for each square-lattice layer. 
The latter is of order the superconducting coherence length at
zero temperature.
Also, $J_{\parallel} = (d/4\pi) (\Phi_0/2\pi\lambda_{\rm L})^2$ is the
local phase rigidity within layers, while $J_{\perp}/a^2$
denotes the local Josephson coupling  energy per unit area.
It is important to observe that the 
Josephson penetration length at zero temperature,
$\gamma_0^{\prime} a$,
provides a natural scale for  the $XY$ model (1)  in the limit of
weak coupling between layers,  in which case
the model anisotropy parameter 
$\gamma_0^{\prime} = (J_{\parallel}/J_{\perp})^{1/2}$ is  much larger
than unity.$^{4}$  This shall be assumed throughout.   Last, we remind
the reader that any generalized phase auto-correlation function
set by an     integer source field, $p(r)$, is related to the
corresponding partition function
$$Z_{XY}^{(3)}[p]=
\int {\cal D} \phi\,  e^{-E_{XY}^{(3)}/k_B T} e^{ i\sum p  \phi}\eqno (2)$$
by the quotient$^{17,18}$
$$\Bigl\langle {\rm exp} \Bigl[i\sum_r p(r) \phi(r)\Bigr]\Bigr\rangle = 
Z_{XY}^{(3)}[p]/Z_{XY}^{(3)}[0].\eqno (3)$$
A knowledge of these amplitudes characterizes the thermodynamics
of the layered $XY$ model (1).

{\it A. Layered Coulomb Gas.}
We shall now employ the well-known dual representation
of the $XY$ model (1) that is based on the Fourier series expansion$^{18}$
$$e^{\beta{\rm cos} \theta} =
 \sum_{n = -\infty}^{\infty} I_{|n|} (\beta) e^{in\theta}$$
 of  the Gibbs
distribution in terms of the modified Bessel functions $I_n (x)$.  
This identity allows the phase variables to be integrated out of the partition
function at the price of introducing a new integer field $n$
that lives on each link of the $XY$ model (1).  
Substitution into Eq. (2) thereby
results      in the dual form
$$Z_{XY}^{(3)}[p]  = I_0^{{\cal N}^{\prime}}(\beta_{\perp})
I_0^{2\cal  N}(\beta_{\parallel})
\sum_{\{n_{\mu}(r)\}}\Pi_{r}
\delta\Bigl[\sum_{\nu}\Delta_{\nu} n_{\nu}|_{r} - p(r)\Bigr]\cdot 
\Pi_{r,\nu} t_{n_{\nu}(r)}(\beta_{\nu})
e^{- i n_{\nu}(r)A_{\nu}(r)}\eqno (4)$$
expressed in terms of modified Bessel functions and the quotients
$t_n(\beta) = I_{|n|}(\beta)/I_0(\beta)$.
Above, $n_{\mu}(r)$ is  an integer link-field
on the layered lattice structure  of
points $r = (\vec r,l)$, with   $\mu = \hat x, \hat y, \hat z$,
and $\beta_{\mu} = J_{\mu}/k_B T$.  Also, $\cal N$ denotes the
total number of sites, while ${\cal N}^{\prime} = {\cal N} (1 - N^{-1})$
gives the total number of rungs between layers.
[The perpendicular link  fields at the boundary layers are set to
$n_z(\vec r, 0) = 0 = n_z(\vec r, N)]$.
We now take the crucial step in the theory by observing that  only
the configurations with $n_z(r) = 0, \pm 1$ are relevant
in the weak-coupling limit,$^{35, 36}$
$\beta_{\perp}\ll 1$.
   After re summation over the parallel link fields
$n_x$ and $n_y$ in Eq. (4), we then obtain the form
$$Z_{XY}^{(3)}[p]  = 
I_0^{{\cal N}^{\prime}}(\beta_{\perp}) \sum_{\{n_{z}(r)\}} y_0^{N[n_z]}\cdot
\Pi_{l=1}^N Z_{XY}^{(2)}[q_l] \cdot e^{-i\sum_r n_z A_z} \eqno (5)$$
for the partition function of the 3D $XY$ model
in terms of sums of  products over partition functions 
$$Z_{XY}^{(2)}[q_l]=
\int {\cal D} \phi_l\,  
e^{-E_{XY}^{(2)}/k_B T} e^{ i\sum q_l  \phi_l}\eqno (6)$$
for  frustrated 2D $XY$ layers in isolation,  
where 
$$E_{XY}^{(2)} = - J_{\parallel}
\sum_{\vec r}\sum_{\mu=x,y}
{\rm cos}[\Delta_{\mu}\phi_l(\vec r)-A_{\mu}(\vec r)]
\eqno (7)$$
is the intra-layer superfluid  kinetic energy, and where
$$q_l(\vec r) =  p(\vec r, l) + n_z(\vec r, l-1) - n_z(\vec r, l)
\eqno (8)$$
is the source integer field.
Above also, $N[n_z] = \sum_{\vec r,l} |n_z(\vec r,l)|$ is the
number of $n_z$ (fluxon) charges per configuration
and 
$y_0 = I_1(\beta_{\perp})/I_0(\beta_{\perp})$
is the bare fugacity.
The latter tends to 
$$y_0 =  \beta_{\perp}/2\eqno (9)$$
in the weak-coupling limit $\beta_{\perp}\ll 1$.
By analogy with Eq. (3), we now identify
the quotient
$Z_{XY}^{(2)}[q_l]/Z_{XY}^{(2)}[0]$ with the
generalized  auto-correlation function
$$C_l [q_l] = \Bigl\langle {\rm exp} \Bigl[ i \sum_{\vec r}
q_l(\vec r) \phi(\vec r,l)\Bigr]\Bigr\rangle_{J_{\perp} = 0}\eqno (10)$$
of the 2D $XY$ model with frustration.$^{17,18}$
Comparison with Eqs. (5) and (6) thus yields the
form
$$Z_{XY}^{(3)}[p]  = I_0^{{\cal N}^{\prime}}(\beta_{\perp}) \cdot
Z_{\rm CG}[p]\cdot \Pi_{l = 1}^N Z_{XY}^{(2)}[0] \eqno (11)$$
for the partition function of the layered $XY$ model
in terms of a product of a  layered  Coulomb gas ensemble
$$Z_{\rm CG}[p] = \sum_{\{n_{z}(r)\}} y_0^{N[n_z]}\cdot
\Pi_{l = 1}^N C_l   [q_l]\cdot
e^{-i\sum_r n_z A_z}   \eqno (12)$$
with $N$ $XY$ model layers in isolation (see  ref. 27).
This is the final result of taking the selective high-temperature
limit, $y_0\rightarrow 0$, which   is assumed throughout.

Before proceeding further, we now derive a useful relation 
between the density of $n_z$ charges and the local Josephson
coupling.  Eq. (1) indicates that the latter is given by
$$\langle {\rm cos}\,\phi_{l,l+1}\rangle = {\cal N}^{\prime - 1}
\partial \,{\rm ln}\, Z_{XY}^{(3)}[0]  /
 \partial \beta_{\perp},\eqno (13)$$
where 
$\phi_{l,l+1}(\vec r) = \phi (\vec r,l+1)- \phi (\vec r,l)-A_z(\vec r)$
is the gauge-invariant phase difference between consecutive layers.
Yet the Coulomb-gas ensemble  (12)  yields the identity
$\langle N[n_z] \rangle = 
y_0 (\partial\, {\rm ln}\, Z_{\rm CG} [0]/\partial y_0)$
for the total number of fluxons on average.
By the factorization 
(11) for the layered $XY$ model, we thereby
obtain the general expression
$$\langle N[n_z]\rangle  /{\cal N}^{\prime}  =
2 y_0 (\langle {\rm cos}\, \phi_{l,l+1}\rangle - y_0)\eqno (14)$$
for this quantity per rung in terms of the local Josephson
coupling.

To demonstrate that (12) is indeed a layered Coulomb gas ensemble, consider
now a single  neutral  pair of unit $n_z$ charges
that lie  in between
layers $l^{\prime}$ and $l^{\prime}+1$ in the absence of an external
source, $p = 0$, with the
negative and positive charges located at planar sites
$\vec r_1$ and $\vec r_2$, respectively.
This represents the first non-trivial term
in the selective  high-temperature expansion (12).
The gauge-invariant product over intra-layer autocorrelation functions 
in the layered Coulomb gas ensemble then reduces
to the product
$C_{l^{\prime}} (\vec r_1,\vec r_2) C_{l^{\prime}+1}^{*}(\vec r_1,\vec r_2)$
of the corresponding  
two-point  functions,
$$C_{l} (\vec r_1, \vec r_2) = \Bigl\langle {\rm exp}
\Bigl[i\phi (\vec r_1,l) - i\phi (\vec r_2,l)
\Bigr]\Bigr\rangle_{J_{\perp} = 0}\ ,
\eqno (15)$$
within isolated layers.
This function takes the form
$$C_{l} (\vec r_1, \vec r_2) = |C_{l} (\vec r_1 -\vec r_2)|
e^{-i\int_1^2 \vec A^{\prime}(\vec r)\cdot d\vec r}\eqno (16)$$
in terms of
a suitably  gauge-transformed vector potential
$\vec  A^{\prime}$, and in terms of
a  magnitude that varies with the separation  as
$$|C_l (\vec r)| = g_0 
(r_0/|\vec r\,|)^{\eta_{2D}} 
\quad {\rm for} \quad |\vec r|\ll \xi_{\rm vx}, \eqno (17)$$
and as
$$|C_l (\vec r)| = g_0 \, {\rm exp}
(-|\vec r\,|/\xi_{\rm vx})
\quad {\rm for}\quad |\vec r|\gg \xi_{\rm vx}.\eqno (18)$$
Here, 
$$\eta_{2D} = \eta_{\rm sw} + \eta_{\rm vx}\eqno (19)$$
is the 2D correlation exponent inside layer $l$,  where 
$\eta_{\rm sw} = (2\pi\beta_{\parallel})^{-1}$
and  $\eta_{\rm vx}$ are the  respective
spin-wave and vortex contributions 
(see Appendix A).
Also,
$\xi_{\rm vx}$ denotes the 2D
phase correlation length, while   the length
$r_0 = a_{\rm vx} / 2^{3/2} e^{\gamma_{\rm E}}$ is  set by
the inter-vortex scale,
$$a_{\rm vx} = (\Phi_0/B_{\perp})^{1/2},\eqno (20)$$
 and by
Euler's constant,$^{18}$ $\gamma_{\rm E}$.
General scaling considerations$^{37}$ indicate
 that the ratio $\xi_{\rm vx}/a_{\rm vx}$ depends only on
temperature.
Last, the prefactor  in expression (17) is of order
$g_0\sim (a/a_{\rm vx})^{\eta_{\rm sw}}.$
 At fields $B_{\perp} > \Phi_0/\gamma_0^{\prime 2} a^2$,
it is therefore  of order unity for 
anisotropy parameters,
$\gamma_0^{\prime} < e^{2\pi\beta_{\parallel}}$,
that are not astronomically large.
The effective layered CG ensemble (12)
therefore takes   form$^{19}$
$$\eqalignno{ Z_{\rm CG} [0] \cong  \sum_{\{n_z\}}  y^{N[n_z]}{\rm exp}
\Biggl\{  - {1\over 2}\sum_{l}
\sum_{\vec r_1 \neq \vec r_2}^{\qquad\prime} q_l(\vec r_1)
\Bigl[ \eta_{2D}
 & {\rm ln} (r_0/|\vec r_1 - \vec r_2|)
 - V^{[q_l]}_{\rm string} (\vec r_1,  \vec r_2)\Bigr]
q_l(\vec r_2) \cr
&-i\sum_{l}\sum_{\vec r}^{\qquad\prime} n_z(\vec r, l) A_z(\vec r,l)\Biggr\}
 &(21)\cr}$$
in the limit of dilute fluxon ($n_z$) charges.
Above, the Coulomb gas ensemble (12)
has been coarse grained up to the natural ultra-violet scale $a_{\rm vx}$.
In particular, the sums above are restricted to a square sublattice
with lattice constant $a_{\rm vx}$.
This  requires the introduction of an effective coarse-grained fugacity
$$y =  g_0 (a_{\rm vx}/a)^{2}
y_0.\eqno (22)$$  
At relatively small separations, 
$|\vec r_1 - \vec r_2| \ll \xi_{\rm vx}$,
within a 2D correlation length, the fluxons experience
a pure ($V_{\rm string}^{[q_l]} = 0$) Coulomb interaction
set by 
$\eta_{2D}$ [see Eq. (17)].
At large separations $|\vec r_1 - \vec r_2|\gg\xi_{\rm vx}$,
on the other hand,
the fluxons experience
a pure ($\eta_{2D} = 0$) confining interaction
$V_{\rm string}^{[q_l]} (\vec r_1,\vec r_2) =
   |\vec r_1  -  \vec r_2|/\xi_{\rm vx}$
[see Eq. (18) and refs. 35, 38 and 39].  
Study of the asymptotic behavior exhibited by general
$n$-point auto-correlation functions (10) for the 2D
$XY$ model reveals that the form (21) for the Coulomb
gas ensemble (12) remains valid in the case
of a general fluxon charge configuration, $n_z(\vec r, l)$
(see Appendix A).  However, the  preceding confining interaction
exists only
between those  well separated points  $\vec r_1$ and $\vec r_2$
in a given layer  $l$
that are connected by a  string. 

The coarse-graining (21) of the original Coulomb gas ensemble (12)
assumes implicitly that there be no more
than one $n_z = \pm 1$ charge per unit sublattice
area $a_{\rm vx}^2$.  This is equivalent to the condition 
$\langle N[n_z]\rangle  /{\cal N}^{\prime} < a^2/a_{\rm vx}^2$
for the density of such fluxons.  Comparison with expression
(14) for this density  then implies that the preceding condition is satisfied
at  temperatures and fields such that 
$\beta_{\perp} a_{\rm vx}^2/a^2 < 1$.  The validity of
the above coarse-grained Coulomb-gas ensemble (21) is  hence assured
at high perpendicular fields
$B_{\perp} > \beta_{\parallel} \Phi_0/\gamma_0^{\prime 2} a^2$.
Below, we determine the thermodynamic 
nature of the two phases that correspond
to quasi long-range  (17) and to short-range (18)
phase correlations within isolated layers.

{\it B. Coupled Phase.}
Consider first  the case (17) where quasi long-range intra-layer
phase correlations are present: $\xi_{\rm vx} = \infty$ and
$V^{[q_l]}_{\rm string} = 0 $.  
Inter-layer $n_z$ charges in the Coulomb gas ensemble are
screened at low temperatures,  $\eta_{2D} < 2$.
Global charge conservation is no longer
enforced in this instance.
Let us also assume  the weak-coupling regime defined
by the inequality $y\ll 1$ to be satisfied by the 
effective fugacity (22) of the coarse-grained
CG ensemble (21).
Following the standard prescription,$^{35,36}$
we then  sum {\it independently} over charge configurations at each site,
with the
restriction  to values $n_z = 0, \pm 1$.  
An appropriate Hubbard-Stratonovich transformation$^{24}$
of the CG (21) in the absence of a source ($p = 0$) followed by
a Debye-H\" uckel type of approximation$^{35, 36}$ reveals that 
it is equal 
to the corresponding one
$$Z_{\rm LD}[0] = \int {\cal D} \theta\,  e^{-E_{\rm LD}/k_B T}\eqno (23)$$
for a   renormalized Lawrence-Doniach (LD) model$^{4}$
up to a factor that depends only on $\beta_{\parallel}$.
The LD energy functional 
$$\eqalignno{
E_{\rm LD} = 
 \bar J_{\parallel}\sum_{l = 1}^{N}\sum_{\mu = x,y}\sum_{\vec r}^{\qquad \prime}
 {1 \over 2}(\Delta_{\mu}^{\prime}\theta_l)^2
 - 2 y k_B T\sum_{l = 1}^{N-1}\sum_{\vec r}^{\qquad \prime}
{\rm cos}(\theta_{l+1} -
\theta_{l} - A_z)\cr
&&(24)\cr}$$
is summed over the coarse-grained sublattice.  Here,
$$\bar J_{\parallel} = k_B T / 2\pi \eta_{2D}\eqno (25)$$
is the  macroscopic  phase rigidity of an isolated layer.$^{40}$
(The validity of the Debye-H\" uckel type approximation
will be established {\it a posteriori}.)
Taking the continuum limit of  the
lattice energy (24) yields the
more familiar expression
$$\eqalignno{
E_{\rm LD} = &
\bar J_{\parallel}\int d^2 r \Biggl[
\sum_{l = 1}^{N}
{1\over 2}(\vec\nabla\theta_l)^2
-\Lambda_0^{-2}
\sum_{l = 1}^{N-1}{\rm cos}(\theta_{l+1}
-\theta_l - A_z)\Biggr]
&(26)\cr}$$
for the energy functional of the LD model,
with a renormalized Josephson penetration length 
$$\Lambda_0 = a (\bar\beta_{\parallel}/2 g_0  y_0)^{1/2},\eqno (27)$$
expressed in terms of the parameter
$\bar\beta_{\parallel} = 
 \bar J_{\parallel}/k_B T$.
Substitution of the bare fugacity (9) that corresponds to the
conventional $XY$ model yields a more familiar expression
$$\Lambda_0 = a (\bar J_{\parallel}/ g_0  J_{\perp})^{1/2}
\eqno (28)$$
for the renormalized Josephson scale.
The above continuum
description (26) is understood to have an  ultraviolet
cut off $\alpha_0\sim a_{\rm vx}$ on the order of the coarse-grained
sub-lattice constant.  It is also {\it gaussian} within layers, 
which means that 
parallel Josephson vortices are the only topological
excitations  that it contains explicitly.

The above renormalized  LD  model is known to be macroscopically
Josephson  coupled at temperatures below
$k_B T_{*0} =     4\pi \bar J_{\parallel}$.$^{21-24,41,42}$
To illustrate this fact,
consider first the minimal double-layer case, $N = 2$.
The zero-temperature line tension of  a single Josephson vortex
is equal to  the linear  energy density
$$\varepsilon_{\parallel}(0) = 
2^{5/2}\bar J_{\parallel}/\Lambda_0\eqno (29)$$
of a sine-Gordon soliton.$^{24}$
The condensation energy, on the other hand, is in general equal to
$-G_{\rm cond} = k_B T\, {\rm ln}\, Z_{\rm LD}$.
At zero temperature  and zero parallel field,  this energy per vertical  rung
is thus given by  
$$-G_{\rm cond}^{(0)} (0)/{\cal N}^{\prime} 
= \bar J_{\parallel} a^2/\Lambda_0^2.\eqno (30)$$
At the unbinding
temperature $k_B T_{\rm LE} = 2\pi\bar J_{\parallel}$ for Josephson
vortex/anti-vortex pairs,$^{41,42}$ on the other  hand,
the double-layer LD model can be analyzed exactly through a mapping
to ideal spinless fermions (see Appendix B).  The line-tension
so obtained is given by the expression
$$\varepsilon_{\parallel} (T_{\rm LE}) = 
\pi\bar J_{\parallel}\alpha_0/\Lambda_0^2,
\eqno (31)$$
where $\alpha_0$
is the natural ultraviolet length scale
of order the  underlying
lattice constant $a_{\rm vx}$ for LD model (26).
This mapping can similarly be employed to obtain the
expression
$$ - G_{\rm cond}^{(0)} (T_{\rm LE}) /{\cal N}^{\prime}  = 
(\bar J_{\parallel}/8) (a \alpha_0/\Lambda_0^2)^2
\eqno (32)$$
for the condensation energy
at zero   parallel field.  Notice that both the line-tension
and the condensation energy above are renormalized down with respect
to their zero-temperature values by factors of $\alpha_0/\Lambda_0$.
Finally, both the fermion analogy$^{42}$  (see Appendix B)
 and the original CG ensemble (21) indicate 
that the renormalized LD model (26) decouples at
$k_B T_* \cong  4\pi\bar J_{\parallel}$.$^{21-24}$  
The fermion analogy in particular
yields    that both the
line tension $\varepsilon_{\parallel}(T)$ and the condensation
energy $G_{\rm cond}^{(0)} (T)$ vanish in an exponentially activated
fashion as $T$ approaches $T_*$ from below (see Appendix B).
A ``semi-classical'' calculation$^{43}$ of
the renormalization  to the zero-temperature
line tension $\varepsilon_{\parallel}$
that results from thermal wandering of  the Josephson vortex 
is consistent with the
exact results listed above at   the
three temperatures 
$T = 0, T_{\rm LE}, \ {\rm and} \ T_*$.$^{24}$  This calculation
obtains a line-tension of the form
$$\varepsilon_{\parallel} = 
4 \bar J_{\parallel}/\lambda_{\rm J},\eqno (33)$$
where $\lambda_{\rm J}$ is a renormalized Josephson penetration length
that varies with temperature following
$${\lambda_{\rm J}\over {r_0}} 
= \Biggl({\Lambda_0\over{2^{1/2} 
r_0}}\Biggr)^
{[1-(\eta_{2D}/2)]^{-1}}.\eqno (34)$$
The exponent above takes the values
$1$, $2$ and $\infty$ at the respective temperatures
$T = 0, T_{\rm LE}, \ {\rm and} \ T_*$.  This agrees with Eqs. (29),
and  (31) if the ultraviolet length scales are identified
by $r_0 = (\pi/8)\alpha_0$. It  is also consistent with
a decoupling transition at $T_*$.  

Given the success of   the 
``semi-classical'' result  [Eqs. (33) and (34)] for the line-tension
in the double-layer case, we propose that the condensation
energy per vertical rung  of the sine-Gordon model [Eq. (26) for $N = 2$]
generally takes the form
$$-G_{\rm cond}^{(0)}/{\cal N}^{\prime}  
= h_0 \bar J_{\parallel} a^2/2\lambda_{\rm J}^2\eqno (35)$$
in zero parallel field,  
where $h_0$ is weakly temperature ($\eta_{2D}$)
dependent and of order unity.
Comparison with Eqs. (30) and (32), for example, yields the assignments
$h_0 (0) = 1$ and $h_0(T_{\rm LE}) = 4/\pi^2$
for this prefactor.
Since the Gibbs free energy of a double-layer is
generally given by the formula
$$G_{\rm cond}  \cong
G_{\rm cond}^{(0)} +
(L_x L_y d) (|B_{\parallel}|/\Phi_0) \varepsilon_{\parallel}\eqno (36)$$
in the limit of vanishingly small parallel field,
$B_{\parallel}$, we obtain the final result
$$-G_{\rm cond}  / {\cal N}^{\prime} \cong  \bar J_{\parallel} 
\Biggl[{1\over 2} 
h_0 (a/\lambda_{\rm J})^2 
- 4(a/\lambda_{\rm J})(a d |B_{\parallel}|/\Phi_0)\Biggr]
\eqno (37)$$
for the condensation energy in such case.
We shall now make use of this result in order to compute the local
Josephson coupling (13).
 Since we have the identity 
$Z_{\rm CG} [0] = Z_{\rm LD} [0]$ up to a factor independent
of $\beta_{\perp}$, the factorization (11) implies that
$$\langle {\rm cos}\,\phi_{l,l+1}\rangle = y_0 +
\partial ( -  G_{\rm cond}/{\cal N}^{\prime} k_B T)/
 \partial \beta_{\perp}. \eqno (38)$$
Substitution of the double-layer result  (37) for the Gibbs free energy
above yields the final expression
$$\langle {\rm cos}\,\phi_{l,l+1}\rangle \cong  y_0
 +  h_0 g_0 \nu \Biggl({2 r_0^{2}\over{\Lambda_0^2}}\Biggr)^{\nu - 1}
- h_1     g_0 \nu 
\Biggl({2^{1/2} r_0\over{\Lambda_0}}\Biggr)^{\nu - 1}
\cdot {|B_{\parallel}|\over{B_{\parallel}^*}}
 \eqno (39)$$
for the local Josephson coupling in the limit
of  small parallel fields, where 
$$\nu = [1 - (\eta_{2D}/2)]^{-1}\eqno (40)$$ 
is the temperature
dependent exponent that appears in Eq. (34), 
where $B_{\parallel}^* = \Phi_0/\Lambda_0 d$ is the crossover field
above which Josephson vortices overlap, and where $h_1 = 2^{3/2}$.
The second term on the right-hand side of Eq. (39) agrees with the result
obtained by Glazman and Koshelev$^{44}$
 for the local Josephson coupling in layered
superconductors at zero field ($r_0\sim a$). 

Consider next the renormalized LD model (26) in the case of an
infinite number 
of layers, $N\rightarrow \infty$ .  The zero-temperature 
condensation energy at $B_{\parallel} = 0$
 is  clearly
equal to that of the previous double-layer case (30).  In addition,
mean-field treatments of the fermion analogy for the
renormalized LD model indicate that the Gibbs free energy
$G_{\rm cond}$ at low parallel fields
 is in general given by the original energy
functional (26),$^{42}$ but with a renormalized Josephson penetration
length $\lambda_{\rm J} (T) > \Lambda_0$.
In the continuum limit, we have for example 
$$G_{\rm cond}
\cong G_{\rm cond}^{(0)}(0)  +  
{\bar J_{\parallel}\over{2 d}} 
\int dx dy dz\Biggl[
\Biggl({\partial\theta\over{\partial x}}\Biggr)^2 +
\Biggl({\partial\theta\over{\partial y}}\Biggr)^2 +
\gamma^{-2}\Biggl({\partial\theta\over{\partial z}} -
{A_z\over d}\Biggr)^2\Biggr], \eqno (41)$$
with the effective mass  anisotropy parameter $\gamma = \lambda_{\rm J}/d$. 
The line-tension of a 
single parallel Josephson vortex in the
present case $N \rightarrow \infty$ 
is then given my the known result$^4$
$$\varepsilon_{\parallel} \cong (\pi \bar J_{\parallel}  /   \lambda_{\rm J})
\,{\rm ln} (\lambda_{\rm L}  /   d).\eqno (42)$$
Here, the resulting logarithmically divergent integral (41)
has been cut off by the London penetration length, $\lambda_{\rm L}$,
which is taken to be  the natural infrared scale.
Also, the condensation energy in the absence of parallel field
is  then given by the zero-temperature result, Eq. (30),
but with the bare Josephson scale $\Lambda_0$ replaced by
$\lambda_{\rm J} (T)$.	
If we now assume that $\lambda_{\rm J} (T)$ has the same
scaling   form$^{44}$
as the analogous scale in the double-layer case, Eq. (34),
then a repetition of the previous  steps
yields an expression
for the local Josephson coupling in the presence of a small parallel
field of the form
$$\langle {\rm cos}\,\phi_{l,l+1}\rangle \cong  
{1\over 2} \beta_{\perp}
 +  f_0  \Biggl({r_0\over{\Lambda_{\rm J}}}\Biggr)^{\eta}
- f_1      
\Biggl({r_0\over{\Lambda_{\rm J}}}\Biggr)^{\eta/2}
\cdot {|B_{\parallel}|\over{B_{\parallel}^*}}.
 \eqno (43)$$
Here $r_0 = a_{\rm vx} / 2^{3/2} e^{\gamma_{\rm E}}$, 
$\Lambda_{\rm J}$ is of order $\Lambda_0$,
and   the prefactor $f_1$
is of order $g_0\, {\rm ln} (\lambda_{\rm L}/d)$.   
This result for the local Josephson coupling  
is corroborated by the observation that
the second-term above coincides with the intra-layer
auto-correlation function (17) evaluated at a separation
$r = \Lambda_{\rm J}$ of order $\Lambda_0 = \gamma_0 \cdot d$ !$^{44}$ 
This suggests that the prefactor and the exponent
for the second term have the limiting values
$\eta\rightarrow \eta_{2D}$ and
$f_0\rightarrow g_0$  at low temperatures, $\eta_{2D}\ll 1$,
which is corroborated by
the double-layer result (39). 
The latter corresponds to the assignments
$\eta = (\eta_{2D}^{-1} - {1\over 2})^{-1}$ 
for the effective exponent,$^{44}$
$\Lambda_{\rm J} = \Lambda_0/2^{1/2}$
for the effective anisotropy scale,  and
$f_i = h_i g_0/(1-{1\over 2} \eta_{2D})$
for the prefactors ($i =0, 1$).
Notice that $f_0(0) = g_0$ and $f_0(T_{\rm LE}) = (8/\pi^2) g_0$ in 
this case, which indicates that $f_0/g_0$ is
close to unity at low temperatures $\eta_{2D}\ll 1$.


We now demonstrate the validity of the Debye-H\" uckel type of
approximation referred to earlier.  In passing from the
layered CG ensemble (21) to the renormalized LD model (24),
the amplitudes
$1 + 2 y\, {\rm cos}(\theta_{l+1} - \theta_{l} - A_z)$ that
appear in the
partition function as a result of the Hubbard-Stratonovich
transformation$^{24}$ are replaced by
${\rm exp} [2 y\, {\rm cos}(\theta_{l+1} - \theta_{l} - A_z)]$
(see refs. 35 and 36).
This is valid as long as Debye screening of the fluxon charges occurs,
which requires many such charges within a screening cloud:
$\pi \lambda_{\rm J}^2 \cdot  
\langle N[n_z] \rangle /{\cal N}^{\prime} a^2 > 1$.  
Study of the previous formulas
in the double-layer case,  Eqs. (14), (34), and (39),
yields the result 
${\pi \over 2} (f_0/g_0) \bar\beta_{\parallel}$
 for the former quantity.  The Debye-H\" uckel type approximation
is therefore valid at low temperatures $k_B T < \bar J_{\parallel}$,
which as we shall see in the next section 
contains the present coupled phase.

Last, we shall compute the perpendicular phase rigidity,
$\rho_s^{\perp} =
 {\partial^2\over{\partial A_z^{\prime 2}}}
 (G_{\rm cond}/{\cal N})|_0$
in the coupled phase.
Here, $A_z^{\prime}$ represents the longitudinal component
of  the vector potential that is assumed to be equal across all of
the perpendicular links.
Periodic boundary conditions in the direction perpendicular to the 
layers are implicit.
First, the duality transformation  (4) yields the general expression
$$\rho_s^{\perp}
= {\cal N}^{-1} 
\Bigl \langle \Bigl [\sum_{\vec r, l} n_z (\vec r, l)\Bigr]^2 \Bigr\rangle
  k_B T\eqno (44)$$
for the perpendicular stiffness in terms of
a fluxon average over the dual ensemble (21).
Yet the $n_z$ charges are screened in  the coupled phase.
Such short-range correlations among the fluxons suggests
the approximation
${\cal N}^{-1}\langle  [\sum_{\vec r, l} n_z (\vec r, l)]^2 \rangle
\cong \langle n_z^2 \rangle$
for the average that appears above.
But   since    the fluxon charges are effectively restricted 
to take values $n_z = 0, \pm 1$
in the selective high-temperature limit,
$y_0\rightarrow 0$, 
we also have approximately    that
$\langle n_z^2 \rangle \cong \langle N[n_z]\rangle/{\cal N}^{\prime}$.
Use of Eq. (14) 
thus  yields the  (mean-field) estimate
$$\rho_s^{\perp} / J_{\perp} \cong
\langle {\rm cos}\, \phi_{l,l+1}\rangle - y_0
\eqno (45)$$
for the perpendicular phase rigidity of the uniformly frustrated
layered $XY$ model in the selective high-temperature limit
of the coupled phase.  Comparison with the previous results 
 for the local Josephson coupling
indicates that the
perpendicular phase rigidity decreases with increasing temperature
and field following the sum of the  last two terms on the
right-hand side of Eq. (43).

In conclusion, a macroscopic Josephson effect   exists at low temperatures,
$\eta_{2D} < 2$, in extremely type-II
layered superconductors  if isolated layers display
quasi-long-range order (17). 
Indeed, the analysis just completed demonstrates that this is the case 
even in the weak-coupling 
 regime $\langle {\rm cos}\,\phi_{l,l+1}\rangle \ll 1$
of the coupled phase  reached 
at high perpendicular fields
$B_{\perp}\gg \Phi_0/\gamma_0^{\prime 2} a^2$.

{\it C. Decoupled Phase.}  Consider next the case (18) in which intra-layer
correlations are short range:  $\xi_{\rm vx} < \infty$.  Although
the coarse-grained form (21) of the Coulomb gas ensemble
 remains valid in this instance, a direct analysis starting from the
original form for the ensemble defined by Eqs. (8) and (12) 
is more expeditious.  Since      
the phase auto-correlation
functions (3) of interest are those that
probe inter-layer couplings,  
the associated source field shall be assumed to
take the form
$$p(\vec r, l) = n^{(0)}_z(\vec r, l-1) - n^{(0)}_z(\vec r, l),\eqno (46)$$
where $n^{(0)}_z (r)$ is a fixed fluxon  ``charge impurity'' field.
Notice that the intra-layer sources (8) can then  be simply
 re-expressed in terms
of the net fluxon charge distribution
$$m_z = n^{(0)}_z + n_z\eqno (47)$$
 as
$$q_l (\vec r) = m_z(\vec r, l-1) - m_z(\vec r, l).\eqno (48)$$
This image of  fluxon charge impurities is   
useful in the calculation
of inter-layer autocorrelators.  In particular, Eqs. (3),
(11) and (12) yield
the identity  
$$\Bigl\langle {\rm exp} \Bigl(i\sum_r
[ p(r) \phi(r) - n^{(0)}_z (r) A_z (r)]\Bigr)\Bigr\rangle =
Z_{\rm CG}^{\prime}[p]/Z_{\rm CG}[0] \eqno (49)$$
between the corresponding gauge-invariant quantity
and the quotient
of      the modified partition function
$$Z_{\rm CG}^{\prime} [p] = \sum_{\{n_{z}(r)\}} y_0^{N[n_z]}\cdot
\Pi_{l = 1}^N C_l   [q_l]\cdot
e^{-i\sum_r m_z A_z}  \eqno (50)$$
with the unmodified one (12) in the absence of a source. 
We shall now apply this to the calculation
of the local
Josephson coupling $\langle e^{i\phi_{l,l+1}}\rangle$, which
corresponds to fixing a unit fluxon ``charge impurity'' at some point
in between layers $l$ and $l+1$.   In the selective high-temperature
limit, $y_0\rightarrow 0$, the configurations that contribute to the
numerator (50) of the quotient  are limited 
 to those  with a net fluxon charge $ - N[n_z] = -1$
in   between layers $l$ and $l+1$ only. Likewise,
the denominator $Z_{\rm CG}[0]$ can be
approximated  by unity in such case.  We thereby
obtain Koshelev's formula$^{10,11}$
$$\langle e^{i\phi_{l,l+1}}\rangle \cong  y_0
\int d^2 r
C_l (0, \vec r) C_{l+1}^*  (0,\vec r)  e^{  - i b_{\parallel} x} /a^2\eqno (51)$$
for the local Josephson coupling in the decoupled phase.
Notice that short-range autocorrelations (18)  yield
a finite integral above, while quasi-long-range autocorrelations (17)
yield a divergent integral for 2D correlation exponents
that satisfy  $\eta_{2D} < 1$.
Last,  Eq. (51) also implies that the first derivative
$\partial \langle e^{i\phi_{l,l+1}}\rangle /
 \partial B_{\parallel}$ of the local Josephson coupling
 is null in   zero parallel field.
This result survives to all orders of the selective high-temperature
expansion in the decoupled phase.
Eq.         (38) in turn implies that the condensation energy
satisfies
$\partial^2 G_{\rm cond} /
 \partial \beta_{\perp} \partial B_{\parallel} = 0$ at
$B_{\parallel} = 0$.
We then  have  a null line tension
for Josephson vortices in the decoupled phase, since the latter  yields  
$\varepsilon_{\parallel} = 0 = \varepsilon_{\parallel}|_{J_{\perp} = 0}$.
In conclusion, the Josephson effect
is absent in the weak-coupling limit of
 extremely type-II layered superconductors when the
intra-layer phase correlations of isolated layers are short range.

We shall now demonstrate that phase auto-correlations in between layers
are short-range in the decoupled phase. 
Since the introduction of  parallel magnetic
field generally reduces inter-layer phase correlation
[see Eq. (55) below], it is
sufficient to consider the zero-field case, $B_{\parallel} = 0$.
By the previous discussion, the gauge-invariant phase autocorrelator
$\langle e^{i\phi_{l,l+n}}\rangle$ in between $n + 1$ adjacent layers
is equivalent to fixing a column made up 
of $n\geq 1$ unit fluxon ``charge impurities''
in between layers $l$ and $l+n$. The lowest order configurations 
that contribute to the numerator (50)
of the CG ensemble
are those with net fluxon charge $- N[n_z] = -n$ distributed evenly
in between layers $l$ and $l+n$ only.
Equations (16), (49) and (50) therefore yield
the expression
$$\langle e^{i\phi_{l,l+n}}\rangle =
 (y_0/a^2)^n\Pi_{l^{\prime} = l}^{l+n}
\Biggl[\int d^2 r_{l^{\prime}}
|C_{l^{\prime}}(\vec r_{l^{\prime}})| \Biggr] 
e^{ i\int_l^{l+n} \vec A^{\prime}(\vec r)\cdot d\vec r}
\delta^{(2)} (\vec r_{l} + \vec r_{l+1} + ... + \vec r_{l+n})
\eqno (52)$$
for this quantity to lowest order in the fugacity.
Substitution of the Fourier representation
$\delta^{(2)}(\vec R) = (2\pi)^{-2} \int d^2 q\, e^{i\vec q\cdot\vec R}$
for the 2D $\delta$-function above in addition to replacing the phase
factor by unity then yields the inequality
$$\langle e^{i\phi_{l,l+n}}\rangle \leq
\Biggl[{\cal C}_0 \int {d^2 q\over{(2\pi)^2}} 
\Biggl({{\cal C}_q\over{{\cal C}_0}}\Biggr)^{n+1}\Biggr]
 (y_0 {\cal C}_0/a^2)^n , \eqno (53)$$
where
$${\cal C}_q = 
\int d^2r |C_{l^{\prime}}(\vec r_{l^{\prime}})|
 e^{i\vec q\cdot\vec r_{l^{\prime}}}\eqno (54)$$
is the Fourier transform of the intra-layer autocorrelator.  The latter
are assumed to be identical for each layer.  
[Equation (53) is an equality in the absence of field, $B_{\perp} = 0$, or
for $n = 1$.]
The prefactor in
brackets above (53) typically decays polynomially with $n$
[see Eq. (56) below]. 
We then  conclude
that the inter-layer autocorrelator 
$\langle e^{i\phi_{l,l+n}}\rangle$
decays at least exponentially with $n$ in the weak-coupling limit
$y_0\rightarrow 0$.  This implies that the
macroscopic  phase rigidity,
$\rho_s^{\perp}$, in the direction perpendicular to the layers is
null in the decoupled phase.  Notice that
such a null result is consistent with expression (44) 
and with the fluxon charge neutrality 
that is characteristic of the decoupled phase.

To get a more concrete idea of the results just obtained  
in the selective
high-temperature limit, $y_0\rightarrow 0$, for the decoupled phase,
we shall now assume the form (16) and (18) for the
short-range intra-layer autocorrelator.  
Substitution into (51)  yields the expression
$$\langle {\rm cos}\,\phi_{l, l+1}\rangle
\cong {\pi\over 2}  
 \Biggl[1 + \Biggl({b_{\parallel}\xi_{\rm vx}\over 2}\Biggr)^2\Biggr]^{-3/2}
 \Biggl({\xi_{\rm vx}\over{a_{\rm vx}}}\Biggr)^2 g_0 y\eqno (55)$$
for the local Josephson coupling in parallel field.  
Notice the anti-cyclotronic $1/B_{\perp}$ dependence
and $1/T$ dependence in the above
expression for the local Josephson coupling that is  generally
expected from Koshelev's formula (51).
Notice also the quadratic 
dependence with parallel field that is consistent with a null
parallel line tension, $\varepsilon_{\parallel} = 0$.
It is interesting to remark that
outside of the 2D critical regime, where $\xi_{\rm vx}\sim a_{\rm vx}$, 
the result (55) coincides
with the form (43) for the local Josephson coupling
in the coupled phase. 
The corresponding exponent $\eta = 2$ occurs precisely
at the depinning temperature, $\eta_{2D} = 1$, for Josephson vortices
in the double-layer case (39)!$^{42}$
Last, the inter-layer autocorrelator satisfies Eq. (53), 
with Fourier components
${\cal C}_q = 2\pi g_0 \xi_{\rm vx}^2/(q^2\xi_{\rm vx}^2 + 1)^{3/2}$
for the short-range correlations (18).  We thereby obtain the
inequality
$$\langle e^{i\phi_{l,l+n}}\rangle \leq
{g_0\over{3n + 1}} \,(2\pi  y \xi_{\rm vx}^2/a_{\rm vx}^2)^n\eqno (56)$$
for the autocorrelator in between $n+1$ layers in the weak-coupling
limit of the decoupled phase.

\bigskip
\bigskip
\centerline{\bf III. Mixed Phase}
\bigskip
We shall now employ the theory developed in the preceding section
for a finite number $N$ of 
Josephson-coupled $XY$ layers (1) with uniform frustration$^{14,15,24}$
to determine the thermodynamic nature 
of the mixed phase in  the corresponding 
extremely type-II layered superconductor. 
The Josephson energy will be assumed to follow the temperature
dependence $J_{\perp}\propto T_{c0} - T$ in the vicinity of
the mean-field critical temperature,$^{45}$
 $T_{c0}$, while the anisotropy
parameter $\gamma_0^{\prime} = (J_{\parallel}/J_{\perp})^{1/2}$
will be assumed to be a constant that is large compared to 
unity.

{\it A. Phase Boundaries.}
Consider first the limit of weak  local Josephson coupling,
$\langle {\rm cos}\, \phi_{l,l+1} \rangle \rightarrow 0$.  
Comparison with the  results (43) and (55) 
indicates that this limit coincides  with the
regime of high temperatures and high fields,
$T \gg T_{\rm J} = J_{\perp}/k_B$ and 
$B_{\perp} \gg B_{\perp}^* = \Phi_0/\Lambda_0^2$.
It is well known that an isolated lattice of 2D vortices (6) melts
 at a temperature 
$$k_B T_m^{(2D)} \cong J_{\parallel}/20, \eqno (57)$$
above which quasi long-range positional correlation
is lost.$^{29-31}$  The transition is driven by the unbinding of
dislocation pairs and it  is second-order.$^{32,33}$
This is reflected by the analysis of the Villain model (Appendix A),
where the elementary excitations are explicitly  interstitial/vacancy
pairs with respect to the triangular vortex lattice that exists  at
zero temperature.$^{12,13}$ 
Suppose now that the inter-layer coupling is weak enough
so that the inequality $T_{\rm J} \ll T_m^{(2D)}$ is satisfied.
(This is consistent with
 a minimum anisotropy parameter  
$\gamma_0^{\prime}$
between four and five.)
Let us also make the plausible assumption that the phase correlations within
an isolated layer inherit the quasi long-range behavior shown by the vortex
positions at low temperature.$^{29, 30}$ 
By the analysis of the previous section, we then conclude that the layers
exhibit a macroscopic Josephson effect
at low temperature $T < T_m^{(2D)}$ following
the renormalized LD model (26), while
they decouple at high temperature $T > T_m^{(2D)}$.
In the limit of high perpendicular fields,
which coincides with the present  weak-coupling limit
$\gamma_0^{\prime}\rightarrow \infty$
at fixed field,
we therefore have a low-temperature phase composed of
$N$ 2D vortex lattices that exhibit a Josephson effect, and
that is separated from a decoupled vortex
liquid phase at high temperatures by
a second-order line at $T = T_m^{(2D)}$.
Hence, neither the Friedel scenario$^{25}$ (decoupled superconducting
layers) nor the ``line-liquid'' state$^{14,24,26}$ (coupled normal layers)
are thermodynamically possible in the weak-coupling limit.

Consider next the weak-coupling regime
at small  yet non-vanishing local Josephson coupling,
$\langle {\rm cos}\, \phi_{l,l+1}\rangle \ll 1$.
It is important  to observe  first that
the selective high-temperature
expansion for the local Josephson coupling in the decoupled 
phase  breaks down
at perpendicular fields of order the scale 
$$B_{\perp}^{\times} = 
g_0^2   \beta_{\parallel} (\xi_{\rm vx}/a_{\rm vx})^2 B_{\perp}^*
\eqno (58)$$
in the absence of parallel field.
The corresponding approximation 
(55) is of order unity in such case, which coincides   roughly with the
identification   of length scales $\xi_{\rm vx}\sim \Lambda_0$. 
The selective high-temperature expansion (43) for the coupled
phase, on the other hand,
breaks down at much lower fields of order $B_{\perp}^*$
for temperatures $T\lsim T_m^{(2D)}$.
For a given temperature inside the decoupled phase,
the confining nature of the  CG (12) is most prominent
at perpendicular fields that are much
larger than $B_{\perp}^{\times}$.
In  particular, 
the string interaction (18)
binds together  dilute fluxon-antifluxon pairs into  stable
dipoles of dimension $\xi_{\rm vx}$ that do not overlap
in this limit.
The   regime is best described physically  by
a decoupled vortex liquid with short-range correlations on the
scale of $\xi_{\rm vx}$.$^{10}$
At  weak-coupling, 
$\langle {\rm cos}\, \phi_{l,l+1}\rangle\ll 1$, 
Eq. (55) indicates that the effective
fugacity $y$ of the CG ensemble (21) is small compared to unity.
Eq. (55) also implies  that
the cross-over temperature $T_{\times}$ for a fixed perpendicular
 field $B_{\perp}$ (58) lies inside of  the 2D
 critical regime ($\xi_{\rm vx}\gg a_{\rm vx}$) in such case.
A similar dilute  CG description (21)  exists for
 the  layered $XY$ model {\it without} frustration, 
in which case coarse-graining is absent,
$a_{\rm vx}\rightarrow a$,
 and the effective
fugacity $y$ is replaced by the bare one, $y_0$.$^{20}$
By analogy, we therefore conclude   that 
a {\it second-order} melting transition
takes place
at a temperature $T_m$ that lies
inside of  the dimensional crossover window
$T_m^{(2D)} < T < T_{\times}$ for a  fixed perpendicular field 
$B_{\perp} \gg B_{\perp}^*$ (see ref. 34).
It is worth pointing out that the bound fluxon pairs  begin
to overlap inside of this regime [see Eq. (62) below].

Before we continue, it is useful first to define a decoupling contour
$$\langle {\rm cos}\, \phi_{l,l+1}\rangle = 
\langle {\rm cos}\, \phi_{l,l+1}\rangle_D \eqno (59)$$
in the $T$-$B_{\perp}$ plane for
fixed parallel field, $B_{\parallel}$.
Here, $\langle {\rm cos}\, \phi_{l,l+1}\rangle_D$ is
a constant less than but of order unity.$^{11, 46}$
Its value will be determined {\it a posteriori} below.
In particular,
the result (43) for the  local Josephson
coupling in the	coupled phase yields ({\it i})
a nearly ``vertical''  contour line at temperatures
of order the Josephson energy,  
$k_B T_{\rm J} = J_{\perp}$,  
in the limit of extremely high perpendicular
fields $B_{\perp}$ compared to  $B_{\perp}^*$, 
and ({\it ii})
a contour  line 
at  higher temperatures $T_{\rm J} \ll T < T_m^{(2D)}$
with  a perpendicular field
that increases exponentially with $T_{*0}/T$ as
$$H_D \sim B_{\perp}^* 
(f_0/\langle {\rm cos}\, \phi_{l,l+1}\rangle_D)^{2/\eta}.
\eqno (60)$$ 
In the decoupled phase
at high temperatures outside of the 2D critical region
 ($\xi_{\rm vx} \sim a_{\rm vx}$),
on the other hand, 
the  result (55) for the local Josephson
coupling indicates that 
the decoupling contour   lies at a   perpendicular field
$$H_{D}\sim   
g_0^2   \beta_{\parallel}
 B_{\perp}^*/\langle {\rm cos}\, \phi_{l,l+1}\rangle_D \eqno (61)$$
that decreases monotonically with temperature as $1/T$.
A similar result for the decoupling transition is  obtained using
the elastic medium description.$^{8,9}$
Since this  high-temperature contour 
should  connect smoothly with
the low-temperature one (60)     in the coupled phase,
the contour line (59)  must {\it cross} the 2D melting line
$T = T_m^{(2D)}$ 
at  a perpendicular  field many times $B_{\perp}^*$.

Suppose now that  the local Josephson coupling,
$\langle {\rm cos}\, \phi_{l,l+1}\rangle$,
begins to approach unity,  
which can be
achieved by either lowering the perpendicular field
or the temperature.  
The selective 
high-temperature expansion in the coupled phase (43)
breaks  down at temperatures and fields 
in the vicinity of the
decoupling contour (59). 
Yet the CG ensemble (21) is screened in the Josephson-coupled
phase  at low temperatures
$T <  T_m^{(2D)}$ for small effective fugacity $y$.
The neutral plasma of fluxons is dilute in this instance
by   Eq. (14).
Since an increase in the fluxon density
with respect to the dilute limit can only increase the effect
of screening, 
then no thermodynamic phase transition is possible as a function of
field ($y$) for $T <  T_m^{(2D)}$.  This general argument is
corroborated by the phase diagram for the neutral 2D Coulomb gas.$^{47}$
The break-down of the selective high-temperature
expansion in the coupled phase that occurs
in the vicinity of the decoupling contour (59)
 must therefore signal a crossover into 
a flux-line lattice regime.

On the other hand, what happens in the decoupled phase
as the perpendicular field is lowered through
the cross-over field $B_{\perp}^{\times}$ at temperatures
and fields that are
beyond the 2D critical regime ($\xi_{\rm vx}\sim a_{\rm vx}$)?
It is useful in this case to
determine  first the     point at which  bound  fluxon-antifluxon
pairs begin to overlap.
Since  the average  fluxon occupation per vertical rung
is equal to $2 a^2/r_s^2$, where $r_s$ denotes
the average spacing in between dipoles, 
and since the second term in expression
(14) for this quantity is negligible
when the local Josephson coupling is of order unity, we then have
$$\xi_{\rm vx}^2/r_s^2 \cong  y_0 (\xi_{\rm vx}^2/a^2)
\langle {\rm cos}\, \phi_{l,l+1}\rangle.\eqno (62)$$
Notice that  the prefactor on the right-hand side above is
of order the high-temperature approximation (55) for the
local Josephson coupling.
Since both the approximate (55)  and
the true value for
$\langle {\rm cos}\, \phi_{l,l+1}\rangle$ are of order unity$^{11, 46}$
 along the decoupling contour (59) at temperatures outside of the
2D critical regime, Eq.  (62) then  indicates that
  the fluxon pairs begin to
overlap ($r_s\sim\xi_{\rm vx}$) in such case. 
The dipoles  disassociate, however, once they overlap
due to the ineffectiveness  of the string (18).
The  system must therefore eventually  experience a (inverted)
 phase transition into a screened CG
above a critical coupling
$\langle {\rm cos}\, \phi_{l,l+1}\rangle_D$ 
less than but of
order unity.$^{11, 46}$ 
Also, the phase transition must be first-order
since  there is no diverging length scale nearby
($\xi_{\rm vx}\sim a_{\rm vx}$).$^{48}$
This is consistent with elastic medium descriptions of the
vortex lattice in layered superconductors,$^9$ which also predict
a  first-order decoupling transition
at a perpendicular field  $H_D$ of order (61). 
Last, continuity with the second-order melting/decoupling
transition that takes place at higher fields implies
that this first-order decoupling line must end
where the former line  begins.  We therefore predict a
critical endpoint at a temperature and field of
order $T_m^{(2D)}$ and many times $B_{\perp}^*$, respectively.


Finally, the layered $XY$ model (1) shows a second-order phase
transition in the absence of frustration at a relatively
large critical temperature $k_B T_c\sim J_{\parallel}$
in comparison to the 2D melting temperature (57).$^{34}$
The above first-order decoupling line must therefore
end in the vicinity of this zero-field critical point
as temperature increases.  Also, the previous duality 
analysis can be repeated in its entirety for the zero-field
case, where the replacement $a_{\rm vx}\rightarrow a$ must
be made globally.$^{20}$  Eqs. (43) and (55) then imply  
that the local Josephson coupling is of order unity at
temperatures between 
$T_{\rm J} =  J_{\perp}/k_B$ 
and
${\rm min}\, [2\pi \bar J_{\parallel}/k_B {\rm ln}\,\gamma_0^{\prime},
\, T_c^{(2D)}]$.
We therefore  have $T_D < T_c$,
which is also consistent with the  second critical endpoint
in the vicinity of $T_c$ at zero field.

The above results are summarized by the phase diagram
shown in Fig. 1.  
The  phenomenology$^{45}$
$J_{\perp} = E_{J0}  (T_{c0}-T)/T_{c0}$
for the Josephson  energy
in the vicinity of the zero-field transition at $T_{c0}$ yields
the linear temperature dependence$^{2,8,9}$
$H_{D} (T) =   \gamma_2^{-2} H_{c2} (T)$
for the first-order decoupling field (61), where
$H_{c2} (T)\sim (\Phi_0/a^2) (T_{c0} - T)/T$
is the mean-field perpendicular
upper-critical field, and where
$\gamma_2\sim
\langle {\rm cos}\, \phi_{l,l+1}\rangle_D^{1/2}\cdot (k_B T_{c0}/E_{J0})^{1/2}$
is an  effective anisotropy parameter.  The selective
high-temperature expansion (55) indicates that this decoupling line
is depressed quadratically by parallel field at temperatures
outside of the 2D critical regime.
Such behavior  is consistent
with anisotropic Ginzburg-Landau theory.$^{49}$
Continuity with the 
 contour (60) in the
coupled phase also indicates that the first-order line 
crosses the second-order melting line, $T = T_m$.
The phase boundary then continues up in field along the latter
second-order line.  
Finally, Eq. (43) implies that
 the decoupling contour (59) continues into the coupled
phase at temperatures $T < T_m$ 
down to the Josephson energy scale $T_{\rm J}$.  
The non-zero line tension (42) for Josephson vortices
that is characteristic of the coupled phase results in a
{\it linear} depression of this contour with parallel field [see Eq. (43)
and ref. 50].
Last,
it is possible that a vestige of the second-order melting
transition found  at high fields $B_{\perp} > B_{\perp}^*$ persists
down into the low-field region in the form of a crossover
(see Fig. 1).
This regime, however, is beyond the scope of the weak-coupling approach
elaborated here.

{\it B. Latent Heat and Josephson Plasma Resonance.}
We shall now  estimate 
the latent heat across the  first-order decoupling line (59).
The perpendicular contribution is equal to
$\Delta E_{\perp}/{\cal N}^{\prime} =
 - J_{\perp}\Delta \langle {\rm cos}\, \phi_{l,l+1}\rangle$. 
The corresponding entropy jump is then equal to
$\Delta S_{\perp} = \Delta E_{\perp}/T$.
The entropy jump per vortex,
 $a_{\rm vx}^2 \Delta S_{\perp}/{\cal N}^{\prime} a^2$,
 at the first-order transition
due to the Josephson coupling is therefore given by
$$\Delta S_{\perp}\ {\rm per}\ {\rm vx} = 
  \beta_{\parallel} (B_{\perp}^*/H_D) 
 (- \Delta\langle {\rm cos}\, \phi_{l,l+1}\rangle) k_B.
\eqno (63)$$
Eq. (61) implies that it is
of order 
$ - \Delta\langle {\rm cos}\, \phi_{l,l+1}\rangle  k_B$.
Last, it is perhaps useful to point out that a jump
$$\Delta\langle N[n_z]\rangle  /{\cal N}^{\prime}  =
\beta_{\perp} \Delta \langle {\rm cos}\, \phi_{l,l+1}\rangle\eqno (64)$$
in the number of inter-layer fluxons per rung at the first-order
decoupling transition is implied by the relation (14).


Apart from playing an important role in the
determination of the phase
diagram discussed above, the local
Josephson coupling between layers can also be directly probed
through $c$-axis Josephson plasma resonance (JPR) experiments.
Theory dictates that the  plasma frequency is given by
$\omega_{pl} = \omega_0 \langle {\rm cos}\,\phi_{l, l+1}\rangle^{1/2}$,
where $\omega_0$ is its zero-field
 value.$^{10,11,46}$  
The local Josephson
coupling is given by 
Eq.  (55) in the vortex liquid phase, 
where $\xi_{\rm vx}\sim a_{\rm vx}$.  
This yields Koshelev's result$^{10,11}$
$$(\omega_{pl}/\omega_0)^2 \cong
 (f_0 \Phi_0/B_{\perp} a^2) (J_{\perp}/2 k_B T)\eqno (65)$$
for  the corresponding  JPR contours
at constant frequency $\omega_{pl} \ll \omega_0$,  
where  $f_0$ is weakly temperature dependent and of order unity
($g_0^2$).
Consider next the coupled phase at perpendicular fields
in the weak-coupling  regime
$a \ll a_{\rm vx}\ll \Lambda_0$.  The second term on the right-hand side of
Eq. (43) for the local Josephson coupling is then dominant.
The vortex contribution to the 2D correlation exponent
(19) can be neglected 
in the low-temperature limit [see Appendix A, Eq. (A8)].
We therefore obtain
the corresponding JPR contours
$$(\omega_{pl}/\omega_0)^2 \cong 
f_0 (B_{\rm J}/B_{\perp})^{T/T_{*0}}\eqno (66)$$
at constant frequency in the coupled phase at low temperatures,
$T_{\rm J} \ll T < T_m^{(2D)}$,
where $B_{\rm J}$ is a constant 
perpendicular field scale of order $B_{\perp}^*$.
Notice then that the perpendicular field along a given
contour diverges {\it exponentially} with $T_{*0}/T$
 in the weak-coupling regime,
$\omega_{pl} \ll \omega_0$ [see also Eq. (60)].

{\it C. Bulk Limit.}
The thermodynamic limit, $N\rightarrow\infty$, of an infinite number
of layers must be taken in order to model extremely type-II
layered  superconductors in bulk.  Since the decoupling
contour (59) is determined primarily by the Josephson
coupling in between adjacent layers, it 
should not change much in the case of an infinite
number of layers.  The first-order decoupling transition
 should therefore  survive the bulk limit.$^{51}$ 
With respect to the question of the survival of the  second order
melting line  in the bulk limit,
it is useful to observe
that the  Coulomb gas ensemble (21)
used here to describe the frustrated $XY$ model 
  also describes the unfrustrated
model if one makes the replacement
$a_{\rm vx}\rightarrow a$ for the natural ultraviolet scale.  Yet
it is obvious that the universality class of the unfrustrated $XY$
model should pass from 2D to 3D as $N\rightarrow\infty$.
Given that both the frustrated and unfrustrated layered $XY$
models have a common
Coulomb gas form (21), then the universality class of the
former should pass from the 2D to the 3D
universality class of the  latter
 as $N\rightarrow\infty$.  In other
words, the second-order melting line
should survive the bulk limit as well.

{\it D.  Bulk Pinning.}
Vortices both in real   superconductors
and  in the $XY$ model (1)
can be pinned, respectively, by defects and by
the underlying cubic lattice 
 at low enough temperatures.
To get an   idea of how pinning
can effect the Josephson coupling, we shall  consider the
optimal case where each c-axis flux line
is {\it fixed} to  some type of a correlated pin.  
The latter could represent a twin boundary, a
columnar track, or a model substrate potential.
If we now return to  the analysis of the CG ensemble (21), 
this means
({\it i}) that the
auto-correlation functions (16)-(18) for each layer in isolation
are all identical, and ({\it ii})
that they resemble the auto-correlation function
of the {\it unfrustrated} $XY$ model on the square
lattice due to the fact that vortex fluctuations are suppressed.  
In particular, they show  quasi-long range order
  ($\xi_{\rm vx} = \infty$) at temperatures below
the vortex/anti-vortex unbinding transition,
$k_B T_c^{(2D)} = {\pi\over 2} J_{\parallel}$,
following Eq. (17).
The phase autocorrelations within isolated layers are short range
($\xi_{\rm vx} < \infty$)
at high temperatures $T > T_c^{(2D)}$, on the other hand,
following Eq. (18). 
The previous analysis holds in general, with the exception that the
Coulomb gas ensemble (21) 
is {\it not} coarse grained.  In particular,
 the replacement
$a_{\rm vx}\rightarrow a$ must be made  globally. 
This yields a {\it field-independent}  result
for the local Josephson coupling (13).
No first-order phase transition is therefore expected.
A second-order decoupling transition, however,
remains at a critical temperature $T_c$ that
should   lie
within the dimensional crossover window
$T_c^{(2D)} < T < T_{\times}$ (see ref. 34).   
By the previous
arguments, its universality class should pass from that of
 the 2D  $XY$ model  without frustration to that of  the 3D
one
in the thermodynamic limit of an infinite
number of layers.
A schematic phase diagram that summarizes the above results
is shown in Fig. 2.  

It must be emphasized that the above results
are  obtained under the assumption of optimum correlated pinning, which
may not be realizable at all temperatures and fields.
Indeed, an isolated 2D vortex lattice within the $XY$
model is pinned only at low temperatures $T < T_{dp}^{(2D)}$
inside of the quasi-ordered phase below the
2D melting temperature $T_{m}^{(2D)}$.$^{29, 30}$  The 2D depinning
temperature, $T_{dp}^{(2D)}$, is null in the zero-field
limit, and  it increases linearly with increasing field.
This depinning line, furthermore,
merges with the 2D melting line   $T = T_{m}^{(2D)}$ at
a field of order $\Phi_0/36 a^2$.  Since the focus 
of this paper is the mixed phase of type-II superconductors, 
which effectively have no substrate potential, we shall
defer the analysis of intra-layer intrinsic pinning effects in
the  $XY$ model (1) to future work.


\bigskip
\bigskip
\centerline{\bf IV. Discussion}
\bigskip
Below, we shall compare the results obtained in the previous sections
for the thermodynamics of the mixed phase in extremely type-II layered
superconductors
with numerical simulations of the frustrated $XY$ model, with 
relevant experimental work on high-temperature
superconductors, and with
the flux-line entanglement  ideas developed by Nelson$^{52}$ and
co-workers.$^{13}$

{\it A. Numerical Simulations.}  Let us first review the Monte Carlo simulation
results that have been obtained recently for the uniformly frustrated
$XY$ model on an isolated square lattice.$^{29}$  Both these and simulations
of the associated lattice Coulomb gas$^{30}$ find a vortex-lattice
 melting transition
at a temperature (57) that is independent of field, in accord with
theoretical expectations.$^{31}$   They also find,
however, that pinning 
due to the square-lattice grid
at commensurate flux densities
 can only be neglected
at large inter-vortex scales 
$$a_{\rm vx} > 6 a.\eqno (67)$$
(See the discussion at the end of section III.D.) 
Monte Carlo simulations of the layered $XY$ model
(1) have also been performed.$^{14}$  
The most extensive simulations find evidence
for a single first-order phase transition that coincides with the
decoupling contour (59).$^{15}$  The results reported in ref. 11,
for example, yield a value for 
$\langle {\rm cos}\, \phi_{l,l+1}\rangle_D$
near $1/2$.  These simulations find no evidence for a critical endpoint
 where the first-order transition
terminates with decreasing temperature, however,
contrary to  the theoretical results obtained 
here (see Fig. 1). 

The cause for the above  discrepancy may have two sources.
First, the coarse-grained Coulomb-gas 
ensemble (21) used here to analyze the coupled
phase at low temperatures may not be valid in regimes where
the 2D vortex lattices are pinned by the $XY$ model grid.
This suggests that comparison of theory with the Monte Carlo
simulations of the layered $XY$ model (1) should be restricted to
low perpendicular fields (67).
Second, the phase diagram shown in Fig. 1 is valid only for anisotropy  
parameters inside of  the range
$$5 < \gamma_0^{\prime} < L/2\pi a \eqno (68)$$
for $XY$ layers of dimensions $L$ by $L$.  
The lower bound guarantees the inequality
$T_{\rm J} < T_m^{(2D)}$, while the upper bound insures that
the system is not effectively decoupled at zero temperature.$^{34}$
Most, if not all  of the Monte Carlo simulations performed to
date have not satisfied conditions (67) and (68) simultaneously.
This indicates that  a
comparison with the present theory is premature.

{\it B. High-$T_c$ Superconductivity.}  The phase diagram shown in Fig. 1
for the layered $XY$ model is
consistent with the nature of the flux-lattice melting 
observed in the 
mixed phase of clean high-temperature
superconductors.$^{1-3}$
The latter
melting transition begins at (or near)
the zero field critical point, but it
terminates in the middle of the $T-H_{\perp}$ plane as
the temperature is  lowered.
The multi-critical point obtained here at $T_0\sim T_m^{(2D)}$ and
$H_{0}\gsim B_{\perp}^*$ is roughly  consistent with this effect.
It must be remarked,    however, that the
location of the  critical endpoint observed experimentally
in high-temperature superconductors is sensitive
to the degree of disorder.$^5$

The critical endpoint for the vortex-lattice
melting line observed in clean high-temperature
superconductors also coincides with  a multi-critical point.
In particular, the flux-line lattice frees itself
from pins
through thermal fluctuations across a vertical
line in the $T - H_{\perp}$ plane that extends up from
the critical endpoint.$^{5,6}$  The combination of
surface barriers plus the second-order vortex-lattice
melting transition at $T = T_m$ that is obtained
here theoretically for extremely type-II layered
superconductors could account for this effect
(see Fig. 1 and ref. 1).  

A so-called {\it second peak} line also extends horizontally
to lower temperatures 
in the $T - H_{\perp}$ plane from the critical endpoint 
in clean high-temperature superconductors.$^{7}$  Bulk pinning
is observed to increase dramatically at perpendicular fields
that lie above this line.   
Recent observations of muon spin resonance ($\mu$SR) in the same systems
have  reached similar conclusions.$^{53}$
It is conceivable that this
effect is related to the dimensional  cross-over transition
 that is predicted here in the coupled phase 
between a flux-line lattice phase at low fields and a smectic solid
phase at high fields
(see Fig. 1).   In particular, point pinning is
relatively more effective in the latter phase 
composed of
weakly coupled  2D vortex lattices.
Note, however, that the cross-over predicted here
for a finite number of layers
in the absence of pins
is possibly broad [see Eq. (60)], whereas the second peak
transition is sharp.$^{5-7}$  
This suggests either 
that a true decoupling phase transition occurs in bulk (see ref. 51),
or that  point pins
are what drive the second-peak phenomenon into
a sharp transition.$^{53}$



{\it C. Topological Defects and Flux-line Entanglement.}
Distinct solid phases of vortex matter have been proposed
in the past.$^4$  In particular, Frey, Nelson and Fisher
have presented arguments for the existence of
a ``supersolid'' phase in the mixed phase
of layered superconductors that is characterized by
highly entangled flux lines that preserve the 2D lattice
structure within layers.$^{13}$  The ``supersolid''  
phase is predicted to occurs whenever the temperature, $T_d$,
above which  interstitial/vacancy
pairs proliferate$^{12}$
lies below the melting temperature, $T_m$, of the
quasi-2D vortex lattice.
The phase composed of Josephson-coupled 2D vortex lattices 
that appears at low temperatures and high fields in the
phase diagram for the
layered $XY$ model displayed in  Fig. 1 evidently has physical characteristics
that coincide with the ``supersolid'' phase.$^{51}$
In particular, the flux arrangement
is ordered within planes yet disordered along the field direction.
The quasi-2D solid phase discussed here
is furthermore 
limited to temperatures inside of the window
$T_{D} < T < T_m$ at relatively high
fields $H_{\perp} \gg B_{\perp}^*$.  
If the (Josephson) decoupling temperature
scale $T_{D}$ is then identified with the scale
$T_d$, at which point vacancy/interstitial pairs begin
to proliferate,$^{12}$ we recover precisely the ``supersolid'' regime
that is discussed in ref. 13. 
Last, we remark that an
interstitial/vacancy pair in a single layer is confined by Josephson
vortices that run between the neighboring layers.$^{12}$
This  indicates that $J_{\perp}$ is the limiting energy scale
for such defects.  This in turn makes the identification 
of $T_D$ with     $T_d$ 
plausible, since $k_B T_D$ is also of order the Josephson scale
$J_{\perp}$ in the high-field limit 
by   Eqs. (43) and (59).

\bigskip
\bigskip
\centerline{\bf V. Concluding Remarks}
\bigskip

A schematic phase diagram  (Fig. 1) for the mixed phase
of extremely type-II layered superconductors has been
proposed on the basis of a weak-coupling duality analysis
of the layered $XY$ model with uniform frustration (1).
The artificial pinning effects that arise from the 2D model
grid in  each layer were effectively removed by a
coarse-graining (21) of the resulting  dual Coulomb gas description (12).
We notably predict that a smectic (super$^{13}$) solid type of
vortex matter exists at high fields and low temperatures,
and that this phase undergoes a second-order melting
transition into  a vortex
liquid phase at higher temperatures.  The solid  phase exhibits
a macroscopic Josephson effect, while the liquid phase does not.
The weak-coupling analysis thus indicates that neither an intermediate
``line-liquid'' phase that exhibits a macroscopic Josephson
effect between normal layers, nor a Friedel scenario 
characterized by decoupled superconducting
layers, are thermodynamically possible in pure 
extremely type-II superconductors.
It has also been argued
that the second-order melting transition converts itself into a first-order
one at lower fields of order many times
the dimensional cross-over scale, $B_{\perp}^*$.
This  potentially accounts for the critical endpoint for the first-order
melting of the flux-line lattice 
that is observed in clean high-temperature
superconductors.$^{1-3}$

Notably absent in this theory are effects due to 
 magnetic screening$^{54}$
and to  disorder caused by point pins.$^{55}$
Both of these effects are present to some degree in
the mixed phase of 
all high-temperature superconductors.$^{5,7}$
It remains 
to be seen       how 
such departures from the pure $XY$ model description (1)
for the mixed phase of  extremely type-II  layered superconductors
may change the conclusions reached above.



The author is grateful for the hospitality of the
Instituto de Ciencia de Materiales de Madrid, where
part of this work was completed, and to Nestor Parga and
Charles Creffield for discussions.  He is also
indebted to Lev Bulaevskii for 
 pressing him    to compute the ``cosine''.

\vfill\eject
\centerline{\bf Appendix A: Villain Model} 
\bigskip 
The duality analysis for the
conventional $XY$ model  presented in section II 
can be repeated in its entirety for the
case the $XY$ model in the Villain form:
$$Z_{\rm V}^{(3)}[p] = \sum_{\{n_{\mu}(r)\}}\Pi_{r}
\delta\Biggl[\sum_{\nu}\Delta_{\nu} n_{\nu}|_{r} - p(r)\Biggr]
{\rm exp}\Biggl[-\sum_r\Biggl({1\over {2\beta_{\parallel}}}
\vec n^2 + {1\over{2\beta_{\perp}}} n_z^2
+ i\sum_{\nu} n_{\nu}A_{\nu}\Biggr)\Biggr].\eqno (A1)$$
In the weak-coupling limit, $\beta_{\perp}\rightarrow 0$,
partial resummation of the type outlined at the beginning of
section II.A  yields the factorized   form
$$Z_{\rm V}^{(3)}[p]  = 
Z_{\rm CG}[p]\cdot \Pi_{l = 1}^N Z_{\rm V}^{(2)}[0] \eqno (A2)$$
for the partition function
in terms of the corresponding 2D partition functions
$$\eqalignno{
Z_{\rm V}^{(2)} [q_l] = \sum_{\{\vec n\}}&\Pi_{\vec r}
\delta[\vec\nabla\cdot\vec n|_{\vec r, l}
- q_l(\vec r)]\times\cr
&\times {\rm exp}\Biggl[-{1\over {2\beta_{\parallel}}}
\sum_{\vec r} \vec n^2(\vec r,l)
-i\sum_{\vec r} \vec n(\vec r,l)
\cdot\vec A (\vec r)\Biggr]  & (A3)\cr}$$
and the Coulomb gas ensemble (12).  In contrast 
with the case of the  conventional
$XY$ model, the fugacity is now
equal to
$$y_0 = e^{-1/2\beta_{\perp}}, \eqno (A4)$$
 while the 2D auto-correlation 
functions (10) are set by the quotients
$$C_l [q_l] = Z_{\rm V}^{(2)} [q_l]/Z_{\rm V}^{(2)} [0].\eqno (A5)$$
Repeating the derivation of the renormalized LD model
(26) results in a renormalized
Josephson penetration length (27) that displays
an artificial temperature dependence 
acquired from the fugacity (A4).  Also, Koshelev's formula
(51) indicates that the local Josephson coupling
acquires the same artificial temperature dependence
in the decoupled phase.  

The above results were initially
derived in ref. 19 by  following an alternate route, where
the spin-wave and  the vortex degrees of freedom
within layers are explicit.
The Villain model displays a complete
factorization of such
degrees of freedom.  This permits the 2D auto-correlation functions
(A5) to be computed analytically in the low-temperature ordered
phase (see refs. 17 and 18).  For example,   the two-point
auto-correlation function (15) for an isolated layer $l$
has the asymptotic form
$$\langle e^{i\phi(\vec r_1, l)} 
e^{-i\phi(\vec r_2, l)}\rangle_{J_{\perp} = 0} 
= g_0 \Biggl ({r_0\over {r_{12}}}\Biggr)^{\eta_{2D}}
e^{-i\int_1^2 \vec A^{\prime}(\vec r)\cdot d\vec r}\eqno (A6)$$
in the ordered phase as a function of the
separation $r_{12} = |\vec r_1 - \vec r_2|$,
where $A^{\prime}$ is a suitably gauge-transformed 
vector potential,
where $\eta_{2D} = \eta_{\rm sw} + \eta_{\rm vx}$
is the 2D correlation exponent, and where
$r_0$  is  of order the natural ultra-violet length scale
$a_{\rm vx}$ set by the spacing between vortices.
The latter is in accord with general scaling considerations.$^{37}$
The separate spin-wave and vortex components of
$\eta_{2D}$ are given by
$$\eta_{\rm sw} = (2\pi\beta_{\parallel})^{-1}\eqno (A7)$$
and by  
$$\eta_{\rm vx} = - {\pi\over 4 a^2} \sum_{\vec r\neq 0}
|\vec r|^2 \langle \delta Q_l (0)\delta Q_l (\vec r)\rangle_{\rm vx},
\eqno (A8)$$
respectively, where 
$\delta Q_l (\vec r) = Q_l (\vec r) - Q_0 (\vec r)$
is the deviation of the intra-layer vorticity $Q_l (\vec r)$
at the dual site $\vec r$ in layer $l$ with respect to 
the flux-line lattice, $Q_0 (\vec r)$, that 
threads each and every layer at zero temperature.
(Surface barriers are assumed.)
The result (A8) is obtained through a cummulant
expansion of the dipole pairs so indicated
over a 2D Coulomb gas (vx) with a uniform background charge density
 $\langle Q_l \rangle /a^2 = b_{\perp} /2\pi a$
and a temperature  of $\eta_{\rm sw}$.  
Last, the prefactor in expression (A6) is of order
$$g_0\sim (a/a_{\rm vx})^{\eta_{\rm sw}}.\eqno (A9)$$
The upper bound $\gamma_0^{\prime} < e^{2\pi\beta_{\parallel}}$
on the anisotropy parameter
guarantees that $g_0$ be of order unity for perpendicular fields 
$B_{\perp} > B_{\perp}^*$.

The methods used to
compute the above two-point function (A6) in the ordered
phase can be extended to compute the asymptotic form of any general
$n$-point autocorrelation function (10).  One thereby obtains the
result
$$|C_l[q_l]| = g_0^{n_+ (l)}
{\rm exp} \Biggl[    {1\over 2}
\sum_{\vec r_1 \neq \vec r_2}\eta_{2D} \cdot q_l(\vec r_1)\,
  {\rm ln} (|\vec r_1 - \vec r_2|/r_0)
q_l(\vec r_2)
\Biggr],\eqno  (A10)$$
where $n_+ (l)$ denotes the number of positive (or negative)
probe charges applied to layer $l$.  Notice that this expression is equal to 
the product of two factors, each  
corresponding to the
spin-wave       and to the vortex           components of
the 2D correlation exponent $\eta_{2D}$.
The vortex factor in the generalized auto-correlation function (10)
is obtained through the cummulant expansion of the associated
Coulomb gas ensemble mentioned above [see Eq. (A8) and  refs. 17 and 18].

With respect to the evaluation of the 2D auto-correlation function in
the disordered phase at high temperatures, 
it is more useful to study the original dual form
(A3) for the corresponding partition function,
$Z_{\rm V}^{(2)} (\vec r_1, \vec r_2)$.  The image
of a string with its ends tied to the
points $\vec r_1$ and $\vec r_2$ at a low dual temperature
$\beta_{\parallel}$ indicates the form
$$Z_{\rm V}^{(2)} (\vec r_1, \vec r_2)\propto 
e^{-\sigma_{\parallel} r_{12}/\beta_{\parallel}}
e^{-i\int_1^2 \vec A(\vec r)\cdot d\vec r}$$ 
for the latter, where $\sigma_{\parallel}$ denotes the string tension
(see refs. 35, 38 and 39).
We thereby obtain the asymptotic result
$$\langle e^{i\phi(\vec r_1, l)}
e^{-i\phi(\vec r_2, l)}\rangle_{J_{\perp} = 0}  =  
g_0 e^{- r_{12}/\xi_{\rm vx}}
e^{-i\int_1^2 \vec A(\vec r)\cdot d\vec r}\eqno (A11)$$
for the auto-correlation function of the 2D Villain model
in the disordered phase, where 
$\xi_{\rm vx} = \beta_{\parallel}/\sigma_{\parallel}$
is the correlation length.  The string image can also
be used to compute general $n$-point autocorrelation functions
(10) in the limit where the probe charges correspond to well
separated dipoles.  The autocorrelator in such case is simply
given by the product of the two-point functions (A11) 
represented by each dipole.

\vfill\eject
\centerline{\bf Appendix B: Fermion Analogy}
\bigskip
We shall now determine the thermodynamic character 
of the renormalized LD model 
[Eqs. (25)-(27)]
in parallel magnetic field 
$B_{l,l+1} = (\Phi_0/2\pi d) b_{l, l+1}$ in 
between layers $l$ and $l+1$.  This model is    
defined     by the energy functional
$$\eqalignno{
E_{\rm LD} = &
\bar J_{\parallel}\int d^2 r \Biggl\{
\sum_{l = 1}^{N}
{1\over 2}(\vec\nabla\theta_l)^2
-\Lambda_0^{-2}
\sum_{l = 1}^{N-1}{\rm cos}[\theta_{l+1} (\vec r)
- \theta_l (\vec r) + b_{l,l+1} x]\Biggr\}
&(B1)\cr}$$
of the real phase variable $\theta_l(\vec r)$.
It is known to be equivalent to 
coupled chains of spinless fermions at zero temperature,
where each chain corresponds to a layer.$^{41, 42}$
(Previous analogies for layered superconductors
in parallel field employed fermions
that live {\it in between} consecutive layers.$^{56, 57}$)
Specifically, the Hamiltonian for the fermion
model   is divided into intra-chain and inter-chain pieces,
$H = H_{\parallel} + H_{\perp}$, with  respectively
$$
H_{\parallel} =  \sum_{l = 1}^N\int dx\Biggl[
v_{\rm F} \Bigl(\Psi_L^{\dag}  i\partial_x \Psi_L
- \Psi_R^{\dag} i\partial_x \Psi_R\Bigr)
 + U_{\parallel}\Psi_L^{\dag}\Psi_R^{\dag}
\Psi_L\Psi_R\Biggr]\eqno (B2)$$
and
$$\eqalignno{
H_{\perp} = & U_{\perp} \sum_{l=1}^{N-1}\int dx
\Bigl[\Psi_L^{\dag}(x,l)\Psi_R^{\dag}(x,l+1)\Psi_L(x,l+1)\Psi_R(x,l)
+ {\rm H.c.}\Bigr]. & (B3)\cr}$$
Above, $\Psi_R(x,l)$ and $\Psi_L(x,l)$ represent
the  field operators
for right ($R$) and left
($L$) moving fermions.
The coordinate along the Josephson vortices, $y$,
is related to the imaginary time
variable $\tau$ of the fermion analogy by
$y = v_{\rm F}^{\prime}\tau$.
Here, the Fermi velocity $v_{\rm F}^{\prime} = 
v_{\rm F}\, {\rm sech} \, 2\phi$
is renormalized by the intra-chain
interaction $U_{\parallel}$,$^{58}$
with  ${\rm tanh}\, 2 \phi = U_{\parallel}/2\pi v_{\rm F}$.
Also, $U_{\perp} > 0$ is a repulsive backscattering
interaction energy$^{58}$
in between chains.
The Gibbs free-energy of the LD model (B1)
with respect to the normal state is then found to be
related to the ground-state energy $E_{\rm F}$ of the fermion analogy by
$$G_{\rm cond}/k_B T = (L_y/v_{\rm F}^{\prime}) 
[E_{\rm F}(U_{\perp}) - E_{\rm F} (0)].
\eqno (B4)$$
The identifications
$$\eqalignno{
b_{l,l+1} = & 2\pi(N_{l+1} - N_{l})/L_x, & (B5)\cr
\eta_{2D} = & 2  e^{2\phi}, & (B6)\cr
\Lambda_0^{-2} = & \alpha_0^{-2} (|U_{\perp}|/\pi v_{\rm F}^{\prime})
\eta_{2D}, 
& (B7)\cr}$$
complete the equivalence between the models, where $N_l/L_x$
gives the fermion density in the  $l^{\rm th}$ chain, where
$\eta_{2D} = k_B T/2\pi\bar J_{\parallel}$ is the 2D correlation exponent,
and where $\alpha_0$  is the natural ultraviolet
length scale of the LD model (B1).    
[Note that the above fermion analogy can directly be shown to
be equivalent to the CG ensemble (21) in the screened phase,
$\xi_{\rm vx} = \infty$, without recourse to the
renormalized LD model (B1) (see ref.  41).]

Consider first the minimal case of $N = 2$ layers.  The 
above analogy
then  reduces to a
Luther-Emery (LE) model for pseudo spin-1/2 fermions,$^{58}$ 
which exhibits pseudo spin-charge
separation.  The right-moving and left-moving
spinless fermions corresponding to the pseudo-spin sector, 
$a_k^{\dag}|0\rangle$ and $b_k^{\dag}|0\rangle$,
are governed by an  ideal Hamiltonian 
along the Luther-Emery line $\eta_{2D} =1$
of the form
$H_{\sigma} = \sum_k [v_F^{\prime} k (a_k^{\dag} a_k - b_k^{\dag} b_k) +
 \Delta_{\sigma} a_k^{\dag}  b_k + \Delta_{\sigma}^* b_k^{\dag}  a_k]$.
They exhibit  a spectrum 
$\varepsilon_{\sigma}^{\pm} (k) = 
\pm (v_{\rm F}^{\prime 2}k^2 + \Delta_{\sigma}^2)^{1/2}$
that is characterized by a
 pseudo spin gap
$$\Delta_{\sigma} = U_{\perp}/2\pi\alpha_0.\eqno (B8)$$
In the absence of parallel field, $b_{1,2}$,
  the hole band ($-$) is filled
and the particle band ($+$) is empty.
Since the pseudo-charge sector is independent of the
coupling between chains, $U_{\perp}$,
the groundstate energy density
relative to the decoupled state, $U_{\perp} = 0$,
is therefore equal to
$$[E_{\rm F} (U_{\perp}) - E_{\rm F} (0)]/L_x
 = -   \Delta_{\sigma}^2/4\pi v_{\rm F}^{\prime}.
\eqno (B9)$$
(The logarithmically divergent contribution above
that  is equal to 
$L_x^{-1} \sum_k \Delta_{\sigma} \langle a_k^{\dag} b_k \rangle$
 has been removed by a   normal-ordering type of  procedure.)
By Eq. (B4), the condensation energy per 
vertical rung at the corresponding LE temperature,
$k_B T_{\rm LE} = 2\pi\bar J_{\parallel}$,  is then equal to
$$ - G_{\rm cond}^{(0)} (T_{\rm LE})/{\cal N}^{\prime} = 
(a \Delta_{\sigma}/v_{\rm F}^{\prime})^2 (\bar J_{\parallel}/2)\eqno (B10)$$
in zero parallel  field.
Yet in the limit of vanishingly small parallel field, 
the Gibbs free energy takes the form
given by Eq. (36),
where $\varepsilon_{\parallel}$ denotes the line tension 
for a single Josephson vortex. 
 It is known that
single-particle excitations within
chains inherit the pseudo spin gap (B8)
because of spin-charge separation.
In addition, Eq. (B5) indicates that 
adding   or removing  a  fermion 
from one of the two chains is equivalent to injecting
a Josephson vortex in between the two superconducting  layers.
The equivalence formula (B4) thus yields the  identity
$$\varepsilon_{\parallel} (T_{\rm LE}) = 
(|\Delta_{\sigma}|/v_{\rm F}^{\prime}) k_B T_{\rm LE} \eqno (B11)$$
in between the line-tension and the pseudo spin gap.  
Comparison with Eqs. (B7) and (B8) 
  yields the final expressions
$$ - G_{\rm cond}^{(0)} (T_{\rm LE})/{\cal N}^{\prime}  =    
(a \alpha_0/\Lambda_0^2)^2 (\bar J_{\parallel}/8)
\eqno (B12)$$
and
$$\varepsilon_{\parallel} (T_{\rm LE}) = 
\pi\bar J_{\parallel}\alpha_0/\Lambda_0^2
\eqno (B13)$$
for the  above free energies in terms of the original model (B1)  parameters.

At  arbitrary number
of layers and at arbitrary temperatures, the fermion analogy can be
 treated in the
mean-field  approximation defined by the charge-density wave (CDW)
order parameter
$\chi_l(x) = \langle\Psi_R^{\dag}(x,l)\Psi_L(x,l)\rangle$
 and the associated gap equation$^{42}$
$\Delta_l = U_{\parallel}\chi_l + U_{\perp}
(\chi_{l+1} +\chi_{l-1})$.  
Standard self-consistent calculation yields a   
single-particle gap 
$$|\Delta_l|  = v_{\rm F} \alpha_1^{-1}
 / {\rm sinh}[2\pi v_{\rm F}/(2 U_{\perp}- U_{\parallel})]
= \Delta_0\eqno (B14)$$
for all chains $l$
that are {\it not} at a boundary ($l=1\ {\rm or}\  N$), and
for couplings that  satisfy   $U_{\perp} > U_{\parallel}/2$.
Here, $\alpha_1$ is the ultraviolet
cut-off for the
mean-field theory, which is  of order $\alpha_0$.
At  temperatures
$T > T_{*0} = 4\pi \bar J_{\parallel}/k_B$ (or $U_{\parallel} > 0$),
we have a null single-particle gap $\Delta_l$ at small couplings
$U_{\perp} < U_{\parallel}/2$ within mean field.
Given the mean-field excitation spectrum
$\varepsilon_l^{\pm} (k) = \pm (v_{\rm F}^2 k^2 +|\Delta_{l}|^2)^{1/2}$
for the fermionic excitations in   chain $l$,
the groundstate energy density
relative to the decoupled state, $U_{\perp} = 0$,
is then equal to
$$[E_{\rm F} (U_{\perp}) - E_{\rm F} (0)]/L_x
 = -   \Delta_{0}^2/4 \pi  v_{\rm F}.
\eqno (B15)$$
By Eq. (B4), the condensation energy per
vertical rung is   equal to
$$ - G_{\rm cond}^{(0)} /{\cal N}^{\prime} =
(a \Delta_{0}/ v_{\rm F})^2 (k_B T/4\pi).\eqno (B16)$$
for temperatures $T > T_{*0}$ in the vicinity of the
decoupling transition.  Eq. (B14) indicates that this quantity
vanishes exponentially at the decoupling temperature  $T_*$
that is  set by
the relationship $U_{\parallel}  = 2 U_{\perp}$.$^{42,58}$


\vfill\eject
\centerline{\bf References}
\vskip 16 pt


\item {1.} G.W. Crabtree and D.R. Nelson, Physics Today {\bf 50}, 38
(April 1997).
 
\item {2.} E. Zeldov, D. Majer, M. Konczykowski,
V.B. Geshkenbein, V.M. Vinokur, and H. Shtrikman,
Nature {\bf 375}, 373 (1995).
 
\item {3.} A. Schilling, R.A. Fisher, N.E. Philips, U. Welp, D. Dasgupta,
W.K. Kwok, and G.W. Crabtree, Nature (London) {\bf 382}, 791 (1996).

\item {4.}  G. Blatter, M.V. Feigel'man, V.B. Geshkenbein, A.I. Larkin,
and V.M. Vinokur, Rev. Mod. Phys. {\rm 66}, 1125 (1994).




\item {5.}  B. Khaykovich, M. Konczykowski, E. Zeldov, R.A. Doyle, D. Majer,
 P.H. Kes, and  T.W. Li, Phys. Rev. B {\bf 56}, R517 (1997).

\item {6.} D.T. Fuchs, E. Zeldov, T. Tamegai, S. Ooi, M. Rappaport
and H. Shtrikman, Phys. Rev. Lett. {\bf 80}, 4971 (1998).

\item {7.} B. Khaykovich, E. Zeldov, D. Majer, T.W. Li, P.H. Kes, and
M. Konczykowski, Phys. Rev. Lett. {\bf 76}, 2555 (1996).

\item {8.} L.I. Glazman and A.E. Koshelev, Phys. Rev. B {\bf 43}, 2835 (1991).

\item {9.} L.L. Daemen, L.N. Bulaevskii, M.P. Maley and
J.Y. Coulter,  Phys. Rev. B {\bf 47}, 11291 (1993).

\item {10.} A.E. Koshelev, Phys. Rev. Lett. {\bf 77}, 3901 (1996).

\item {11.} A.E. Koshelev, Phys. Rev. B {\bf 56}, 11201 (1997).

\item {12.} M. Feigel'man, V.B. Geshkenbein, and A.I. Larkin,
Physica C {\bf 167}, 177 (1990).

\item {13.} E. Frey, D.R. Nelson, and D.S. Fisher,
Phys. Rev. B {\bf 49}, 9723 (1994).

\item {14.} T. Chen and S. Teitel, Phys. Rev. B {\bf 55}, 11766 (1997).

\item {15.} X. Hu, S. Miyashita, and M. Tachiki, Phys. Rev. Lett.
{\bf 79}, 3498 (1997); 
A.K. Nguyen and A. Sudb\o, Phys. Rev. B {\bf 57}, 3123 (1998);
P. Olsson and S. Teitel, Phys. Rev. Lett. {\bf 82}, 2183 (1999).

\item {16.} R. Sasik and D. Stroud, Phys. Rev. Lett. {\bf 75}, 2582 (1995);
J. Hu and A.H. MacDonald, Phys. Rev. B {\bf 56}, 2788 (1997).

\item {17.} J.V. Jos\' e, L.P. Kadanoff, S. Kirkpatrick and
D.R. Nelson, Phys. Rev. B {\bf 16}, 1217 (1977).

\item {18.} C. Itzykson and J.  Drouffe, {\it Statistical Field Theory},
 vol. 1, (Cambridge Univ.  Press, Cambridge, 1991) chap. 4.

\item {19.} J.P. Rodriguez, ``Nature of Decoupling in the Mixed Phase
of Extremely Type-II Layered Superconductors'' (cond-mat/9906199).

\item {20.} J.P. Rodriguez, Physica C {\bf 332}, 343 (2000).

\item {21.}  S.E. Korshunov, Europhys. Lett. {\bf 11}, 757 (1990).

\item {22.} S.E. Korshunov and A.I. Larkin, Phys. Rev. B {\bf 46}, 6395
(1992).

\item {23.} J.P. Rodriguez, Europhys.  Lett. {\bf 31}, 479 (1995).

\item {24.} J.P. Rodriguez, J. Phys. Cond. Matter {\bf 9}, 5117 (1997).

\item {25.} J. Friedel, J. Phys. (Paris) {\bf 49}, 1561 (1988).

\item {26.} Z. Tesanovic, Phys. Rev. B {\bf 59}, 6449 (1999).

\item {27.} N. Parga and J.E. Van Himbergen, Ann. Phys. (N.Y.)
 {\bf 134}, 286 (1981).  
[The Coulomb-gas ensemble (12) 
can be recovered from the
``hybrid'' one
that is derived  in this ref. by    integrating out
the vortex degrees of freedom within layers.]

\item {28.} M. Dzierzawa, M. Zamora, D. Baeriswyl and X. Bagnoud,
Phys. Rev. Lett. {\bf 77}, 3897 (1996).

\item {29.} S.A. Hattel and J.M. Wheatley,
Phys. Rev. B {\bf 51}, 11951 (1995).

\item {30.} M. Franz and S. Teitel, Phys. Rev. Lett. {\bf 73}, 480 (1994);
S. Hattel and J. Wheatley, Phys. Rev. B {\bf 50}, 16590 (1994);
C. E. Creffield and J.P. Rodriguez, unpublished.

\item {31.} B.A. Huberman and S. Doniach, Phys. Rev. Lett. {\bf 43},
950 (1979); D.S. Fisher, Phys. Rev. B {\bf 22}, 1190 (1980).

\item {32.} D.R. Nelson and B.I. Halperin, Phys. Rev. B {\bf 19},
2457 (1979).

\item {33.} J.M. Kosterlitz and D.J. Thouless,
J. Phys. C {\bf 6}, 1181 (1973).

\item {34.} S. Hikami and T. Tsuneto, Prog. Theor. Phys. {\bf 63}, 387 (1980);
see also S.R. Shenoy and B. Chattopadhyay, Phys. Rev. B {\bf 51}, 9129
(1995).

\item {35.} A.M. Polyakov,
{\it Gauge Fields and Strings} (Harwood, New York, 1987).

\item {36.} A.M. Polyakov, Nucl. Phys. B {\bf 120}, 429 (1977).

\item {37.} J.M. Caillol, D. Levesque, J.J. Weis and
J.P. Hansen,  J. Stat. Phys. {\bf 28}, 325 (1982);
S.W. de Leeuw and J.W. Perram, Physica (Amsterdam) {\bf 113A}, 546 (1982).

\item {38.} A. Polyakov, Phys. Lett. {\bf 72} B, 477 (1978).

\item {39.}  J.P. Rodriguez, Phys. Rev. B {\bf 51}, 9348 (1995);
 Phys. Rev. B {\bf 60}, 3689 (E) (1999).

\item {40.} D.R. Nelson and J.M. Kosterlitz, Phys. Rev. Lett. {\bf 39},
1201 (1977).
 
\item {41.} J.P. Rodriguez, Europhys. Lett. {\bf 39}, 195 (1997);
 Europhys. Lett. {\bf 47}, 745 (E) (1999).

\item {42.} J.P. Rodriguez, Phys. Rev. B {\bf 58}, 944 (1998).

\item {43.} R.F. Dashen, B. Hasslacher, and Andr\' e Neveu,
Phys. Rev. D {\bf 11}, 3424 (1975).

\item {44.} L.I. Glazman and A.E. Koshelev, Zh. Eksp. Teor. Fiz. 
 {\bf 97}, 1371 (1990) [Sov. Phys. JETP {\bf 70}, 774 (1990)].

\item {45.} V. Ambegaokar and A. Baratoff, Phys. Rev. Lett. {\bf 10}, 486
(1963); {\bf 11}, 104 (E) (1963).


\item {46.} E.B. Sonin, Phys. Rev. Lett.   {\bf 79}, 3732 (1997).

\item {47.} J.M. Kosterlitz, J. Phys. C {\bf 7}, 1046 (1974).

\item {48.} For a demonstration of a first-order phase transition in the
unconfined 2D Coulomb gas, see
J. Lee and S. Teitel, Phys. Rev. Lett. {\bf 66}, 2100 (1991).
 
\item {49.} G. Blatter, V.B. Geshkenbein and A.I. Larkin, 
Phys. Rev. Lett. {\bf 68}, 875 (1992).

\item {50.} A.E. Koshelev, Phys. Rev. Lett. {\bf 83}, 187 (1999).


\item {51.} Although the Coulomb gas analysis 
[Eqs. (21) and (60)] indicates a decoupling
crossover transition at low temperatures 
$T < T_m^{(2D)}$ for the
case of a finite number of layers, $N$, 
the boson analogy used in ref. 13 predicts,
on the other hand, a true decoupling/super-solid phase 
transition in the thermodynamic limit $N\rightarrow\infty$.  
The compatibility of these two results is assured if the
solid to supersolid transition represents a quantum critical 
point within the boson
analogy.

\item {52.} D.R. Nelson, in {\it Phenomenology 
and Applications of High-Temperature
Superconductors}, Proceedings of the Los Alamos 
Symposium - 1991, edited by
K.S. Bedell, M. Inui, D. Meltzer, J.R. Schrieffer and S. Doniach
(Addison-Wesley, Reading, MA). 
 
\item {53.} J.E. Sonier, J.H. Brewer, R.F. Kiefl, D.A. Bonn,
J. Chakhalian, S.R. Dunsiger, W.N. Hardy, R. Liang,
W.A. MacFarlane, R.I. Miller, D.R. Noakes, T.M. Riseman
and C.E. Stronach, Phys. Rev. B {\bf 61}, R890 (2000). 

\item {54.} G. Blatter, V. Geshkenbein, A. Larkin and H. Nordborg,
Phys. Rev. B {\bf 54}, 72 (1996).

\item {55.} A.E. Koshelev, L.I. Glazman and A.I. Larkin, 
Phys. Rev. B {\bf 53}, 2786 (1996).

 
\item {56.} L.V. Mikheev and E.B. Kolomeisky, Phys. Rev. B {\bf 43},
10431 (1991).

\item {57.} B. Horovitz, Phys. Rev. B {\bf 47}, 5964 (1993).

\item {58.} A. Luther and V.J. Emery, Phys. Rev. Lett. {\bf 33}, 589 (1974);
S. T. Chui and P. A. Lee, Phys. Rev. Lett. {\bf 35}, 315 (1975).











\vfill\eject
\centerline{\bf Figure Captions}
\vskip 20pt
\item {Fig. 1.}   Displayed is a schematic phase diagram for a finite
number of weakly coupled $XY$ layers with uniform frustration [Eq. (1)].
All artificial pinning effects due to the 2D model grids are suppressed
(see refs. 29 and 30).  
The decoupling (dashed/solid) line is a few 
to many times the dimensional crossover
scale $\Phi_0/\gamma_0^{\prime 2} a^2$ at temperatures in the
vicinity of $T_m^{(2D)}$ [see Eqs. (60) and (61)].
Also, the Josephson temperature $T_{\rm J} = J_{\perp}/k_B$ is
assumed to be smaller than the scale of the figure.
Last, the solid phase should be defective on scales that are small
in comparison to the Josephson penetration length $\Lambda_0$
at elevated temperatures $T > T_m^{(2D)}$.

\bigskip

\item {Fig. 2.} Shown is a schematic phase diagram for the layered
$XY$ model (1) under the assumption that all of the vortex lines
are fixed to some type of correlated pin.
Vortex/anti-vortex pairs should begin to proliferate
at elevated temperatures $T_c^{(2D)} < T < T_c$ inside
of the ordered phase on scales that are small in comparison
to the Josephson penetration length $\Lambda_0$.

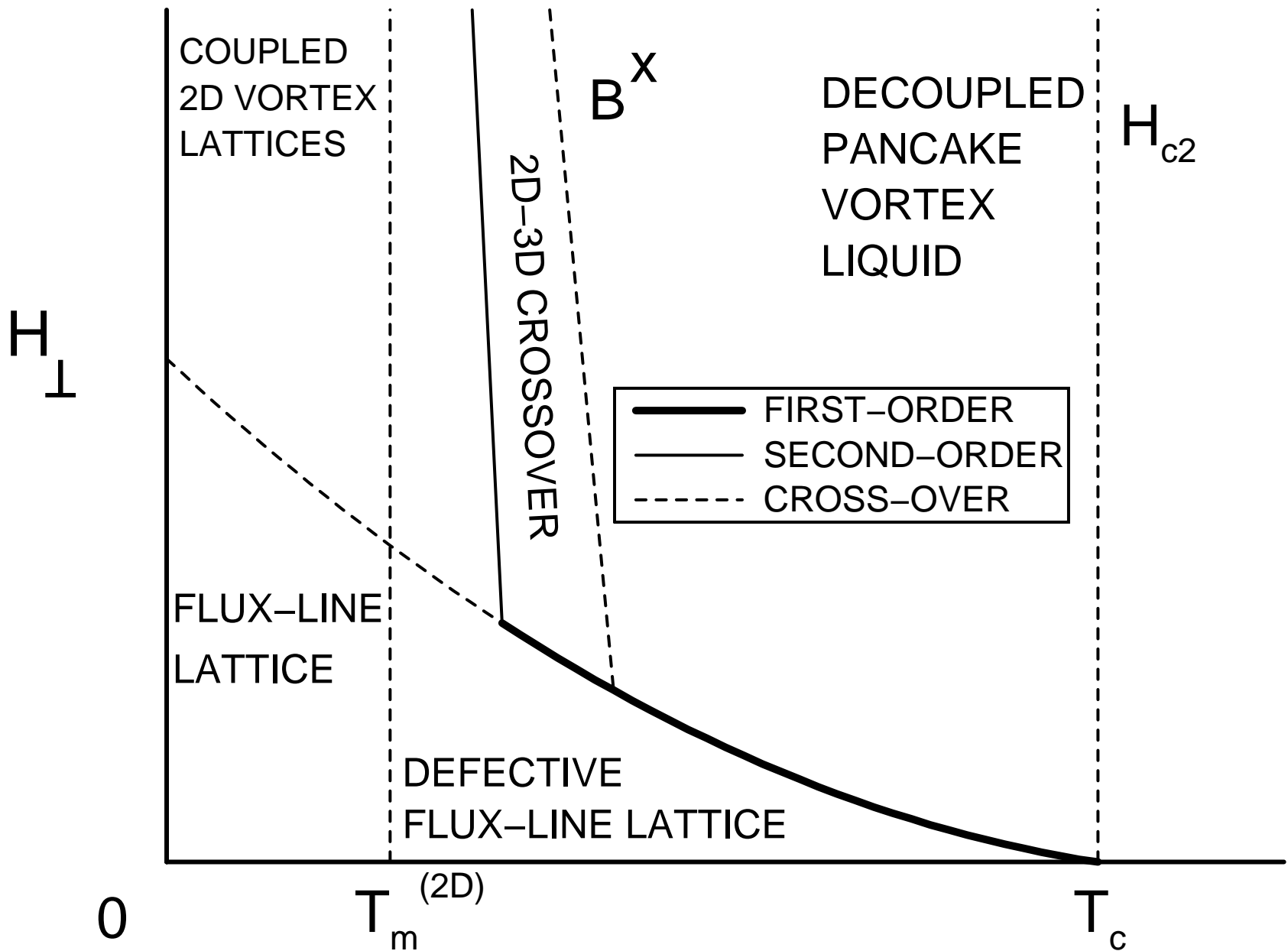
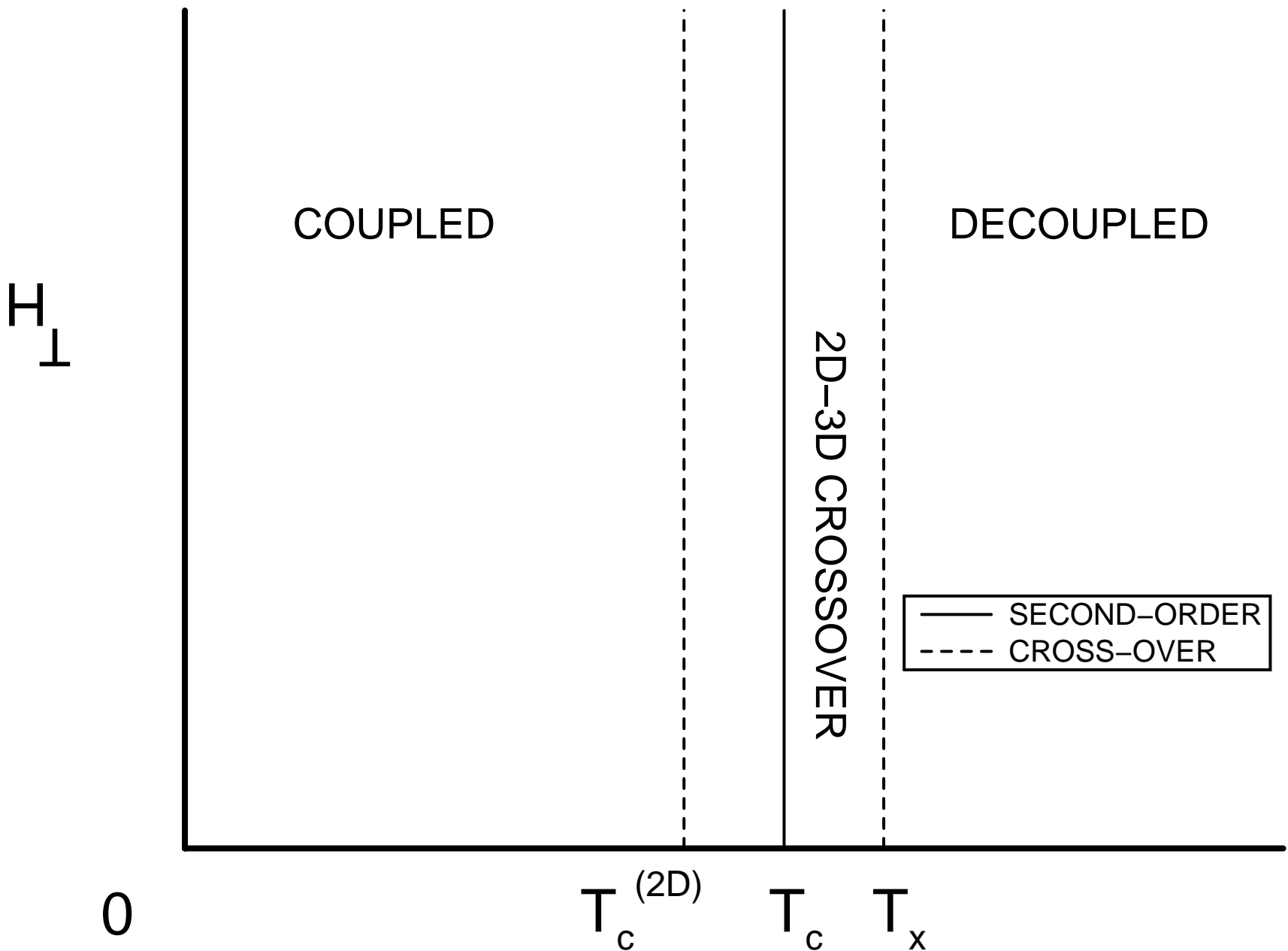
\end